\documentclass[
aps,%
10pt,%
final,%
notitlepage,%
oneside,%
onecolumn,%
nobibnotes,%
nofootinbib,%
superscriptaddress,%
showpacs,%
centertags,%
showkeys,%
amssymb,%
]{revtex4}

  \usepackage[utf8]{inputenc}
  \usepackage{microtype} 
  \usepackage{graphicx}
    \graphicspath{{/}{figures/}}
	\DeclareGraphicsExtensions{.pdf,.png}
    \setkeys{Gin}{width=\linewidth,totalheight=\textheight,keepaspectratio} 
  \usepackage[caption=false]{subfig}
  \usepackage{xcolor}
  \usepackage{comment}
  \usepackage{amsmath,amssymb}
  \usepackage{tabularx,booktabs}
  \usepackage{xspace}

\newcommand*\doctitle{Scientific paper}
\newcommand*\docauthor{Erik Bartoš}

\usepackage[pdftex]{hyperref}
\hypersetup{
      breaklinks    = true,
      baseurl       = http://,
      pdfborder     = 0 0 0,
      pdfpagemode   = UseNone,
      pdfstartpage  = 1,
      pdfcreator    = Pdf\LaTeX{} with hyperref package,
      pdfproducer   = Pdf\LaTeX{},
      bookmarksopen = true,
      pdfauthor     = {\docauthor},
      pdftitle      = {\doctitle},
      pdfsubject    = {Scientific paper},
      pdfkeywords   = {OANDA, string, model, time series, forecasting},
      unicode       = true,
      colorlinks    = true
}

\newcommand{\m}[1]{\mathrm {#1}}
\newcommand{\nn}{\nonumber}

\begin{document}


\title{With string model to time series forecasting}

\author{Richard Pinčák}
 \email{pincak@saske.sk}
\affiliation{Institute of Experimental Physics, Slovak Academy of Sciences,\\ Watsonova 47, 043 53 Košice, Slovak Republic}
\author{Erik Bartoš}%
 \email{erik.bartos@savba.sk}
\affiliation{%
Institute of Physics, Slovak Academy of Sciences, Dúbravská cesta 9, 845 11 Bratislava, Slovak Republic
}%

\date{\today}

\begin{abstract}
Overwhelming majority of econometric models applied on a long term basis in the financial forex market do not work sufficiently well. The reason is that transaction costs and arbitrage opportunity are not included, as this does not simulate the real financial markets. Analyses are not conducted on the non equidistant date but rather on the aggregate date, which is also not a real financial case. In this paper, we would like to show a new way how to analyze and, moreover, forecast financial market. We utilize the projections of the real exchange rate dynamics onto the string-like topology in the OANDA market. The latter approach allows us to build the stable prediction models in trading in the financial forex market. The real application of the multi-string structures is provided to demonstrate our ideas for the solution of the problem of the robust portfolio selection. The comparison with the trend following strategies was performed, the stability of the algorithm on the transaction costs for long trade periods was confirmed.
\end{abstract}

\maketitle

\section{Introduction}\label{sec:introduction}

It has often been argued and there is corroborating empirical evidence which suggests that stock market prices exhibit wave like properties~\cite{A0}. In classical financial mathematics there were conducted fundamental investigations~\cite{R:1,R:1A} to find adequate stochastic processes matching the real financial data: Brownian~\cite{R:2}, geometric Brownian~\cite{R:3}, general Lévy processes~\cite{R:4}. From the point of view of quantum-like approach~\cite{R:5}, the problem cannot even be formulated in such a way, due to the lack of the description of the whole financial market as one single space \cite{FL}. The financial markets have their own intrinsic geometrical structure \cite{R:6,R:7}, but there is no classical stochastic process which would match with the real financial data. Despite some early studies to characterize the nature of topological space of time series data by identifying the state of a financial market \cite{JH, MM} as a complex non-stationary system, there exists no definition of the space of time series data which satisfies the definition of topological space with the fixed-point property, i.~e., it fulfills the $T_{0}$ -- Kolmogorov separation axiom \cite{ZK}. In other words one financial time series are composed of infinitely many random variables in which the average of all of this random variables not always converge to single variable (single Kolmogorov space). Moreover, the fluctuations in the real market are not Gaussian \cite{liz} unless they decay as fast as Gaussian, and the history shows that the catastrophic loss (gain) is more likely than the Gaussian model.

The instantaneous return on the Financial Times-Stock Exchange (FTSE) and all Share Index is viewed as a friction-less particle moving in one-dimensional square well (potential barrier) but where there is a non-trivial probability of the particle tunneling into the well’s retaining walls with some prices uncertainty~\cite{A,B,C,D}. The potential barrier approach is frequently used for barrier option calculations in the stock market~\cite{Jana}.

Combinatorial auctions on financial game theory seem to be the most promising field. The promising quantum-like experiments give rise to commercial implementation of quantum auctions in the near future~\cite{F}. By including a sufficient amount of hidden variables, stochastic models may be able to reproduce the historic observations with similar accuracy. Indeed, the (possible) existence of hidden variables was at the heart of the early critique of quantum theory. It is almost self-evident that the return on a stock depends on many factors that have not been modeled. The question nevertheless is not just one of having a more efficient description of the dynamics. Unlike hidden variables describing physical phenomena, the factors that influence the dynamics of a stock are expected to change over time. Moreover, it is not clear how economic factors such as gross national product influence the value of any given stock at any given point in time. The observed scaling of the return distributions for various stocks in different economic environments strongly suggests that all these ``hidden'' factors find their expression in the mean return and the variance of the returns.

In the economics a problem of the asymmetric information of inefficiency of efficiency market hypothesis still exists. This problem was solved by some micro economists by using the duality theory of Walrasian utility function~\cite{hall} but it is still present in the macro economics time series as so called memory process in the time series of stock price.

The dynamics that determines the shape of the return distributions, on the other hand, must be self-consistent and largely immune to the influence of ``hidden'' variables that are specific for a company and economic and political climate. Hidden variables in our model is an analogy with hidden dimensions in string theory, both are non-visible and non-measurable, however, they have great influence on the whole system.

Especially for the learning of buying and selling signals for the currency pairs as well in financial trading, the position is a binding commitment to buy or sell a given amount of financial instruments. Open positions remain subject to fluctuations in the exchange rate. Open positions are closed by entering into a trade that takes the opposite position to the original trade. The net effect is to bring the total amount for currency pair back to zero. The bid price ($p_{bid}(\tau )$) is always less than the ask price ($p_{ask}(\tau)$) because brokers pay less than they receive for the same currency pair. The spread represents your cost to trade with a broker. The currency pair $p(\tau)$ indicates how much of the quote currency is required to purchase one unit of the base currency, particular currency, which comprises the physical aspects of a nations money supply.

In our string theory approach the hidden variables are $l_s$ -- the length of momentum string, $Q$ -- the quotient or the exponent of the momentum, $m$ -- the frequency of the momentum function and will be described in the next sections.

Almost all known econometric models applied on a long term basis in the financial forex market do not work sufficiently well, as was shown in \cite{P:1}.  The main goal of this paper is the practical demonstration of the usability of the previously derived model in the real market conditions, e.~g., online trading with real data. For this purpose we have used the developing version of the online trade system \cite{Trade} with tick-data level accuracy of simulations. The OANDA database\cite{Oanda} has served as a source of the data, apart from the previous forex market data. As the complex multi-string structures produced by the generalized derivatives of strings cannot be easily grasped by the intuitive principles we have tried to provide the real application of our string approach with a lot of illustrative examples. The proposed string model algorithm behaviour and the stability on the transaction costs was compared with the well known prediction models and trading strategies \cite{ITS,MACD,Arima,Arima2} which serve as the benchmark tests of the more complex time series forecasting models.

In Section \ref{sec:method}, the algorithm of the trade system, the prediction model and the main motivation are presented. In Section \ref{sec:model}, the analyses of the model with a more detailed description and the results are commented. The prediction of long-run profit by means of a spin of strings is sketched in Section \ref{sec:spin}. In the last Section \ref{sec:con}, the conclusions are summarized.

\section{Method}\label{sec:method}

The results of our paper are based on the previous work, where the prediction model based on the deviations from the closed string/pattern form (PMBCS) was derived. The model was tested on the forex market historical data with some interesting results which have confirmed our string theory approach. Here we present only the basic facts important for further understanding, the details and the general overview of the problems are given in \cite{P:1,P:2,P:3}.

All analyses were done in order to maintain a positive value of the net asset value (NAV) for the longest period of time. For all simulations the ask-bid data spreads are included into consideration. It means that all presented results for NAV include also the transaction costs induced by data spreads \cite{W,R}, if not mentioned otherwise. Other types of transaction costs, e.~g., brokers' commissions are not taken into account. The build-in algorithm of the NAV evaluation is as similar as possible to the OANDA algorithm, it was tested on the real data and the deviations were about $0.5\,\%$ only.

\subsection{PMBCS prediction model}\label{subsec:prediction}

In the PMBCS model the incoming currency rates data are transformed into the one dimensional objects ``strings''; mathematically, they are the scalar functions of several variables, which are represented by the momentum of the string.

We have defined the momentum of the string as
\begin{equation} \label{eq:momentum}
M_{(l_s, m, Q, F_{\m{CS}})} =\left(\frac{1}{l_s+1}
\sum_{h=0}^{l_s}\Big| p_{\m{stand}}(\tau,h,l_s) - F_{\m{CS}}(h,l_s)\Big|^Q\right)^{1/Q}
\end{equation}
where
\begin{align}
p_{\m{stand}}(\tau,h,l_s) &= \frac{p(\tau+h) - p_{\min}(\tau,l_s)}{p_{\max}(\tau,l_s)- p_{\min}(\tau,l_s)}\,, \qquad  p_{\m{stand}} \in
(0,1),\\
p_{\max}(\tau,l_s) &= \max_{h\in  \{0,1,2,  \ldots, l_{s}\}}
p(\tau + h) \,,\qquad
p_{\min}(\tau,l_s) = \min_{h \in \{0,1,2,
\ldots, l_s\}} p(\tau+ h), \nn
\end{align}
$h$ denotes a tick lag between currency quotes $p(\tau)$ and $p(\tau+h)$, $\tau$ is the index of the quote. $F_{\m{CS}}(h,l_s)$ is the regular function
\begin{align}
F_{\m{CS}}(h,l_s) = \frac{1}{2} \big(1+\cos(\tilde{\varphi})\big),\quad \tilde\varphi = \frac{2\pi m h}{l_s+1}+\varphi, \label{eq:function}
\end{align}
$\varphi$ is a phase of a periodic function $\cos(\tilde{\varphi})$. The periodic function $\cos(\tilde{\varphi})$ in the definition of the regular function, Eq.~(\ref{eq:function}), could be substituted by different types of mathematical functions.

The momentum defined above takes the values from the interval $M_{(l_s,m,Q,\varphi)}\in(0,1)$. In Sec.~\ref{sec:model}, the behaviour of the momentum with some explicit examples of the regular function $F_{\m{CS}}$ is discussed.

\subsection{Trading algorithm}\label{subsec:strategy}

The main purpose of our work was to demonstrate the usability of the PMBCS model under the different circumstances, as in the previous cases. The results were expected to be performed on the OANDA data with the most realistic trade conditions. For this purpose, the developing version of the trade online system was used \cite{Trade}. It provides highly professional algorithmization of the trading strategies with tick-data level accuracy of simulations.

\begin{figure}[!tb]
\begin{center}
\includegraphics[width=0.39\columnwidth]{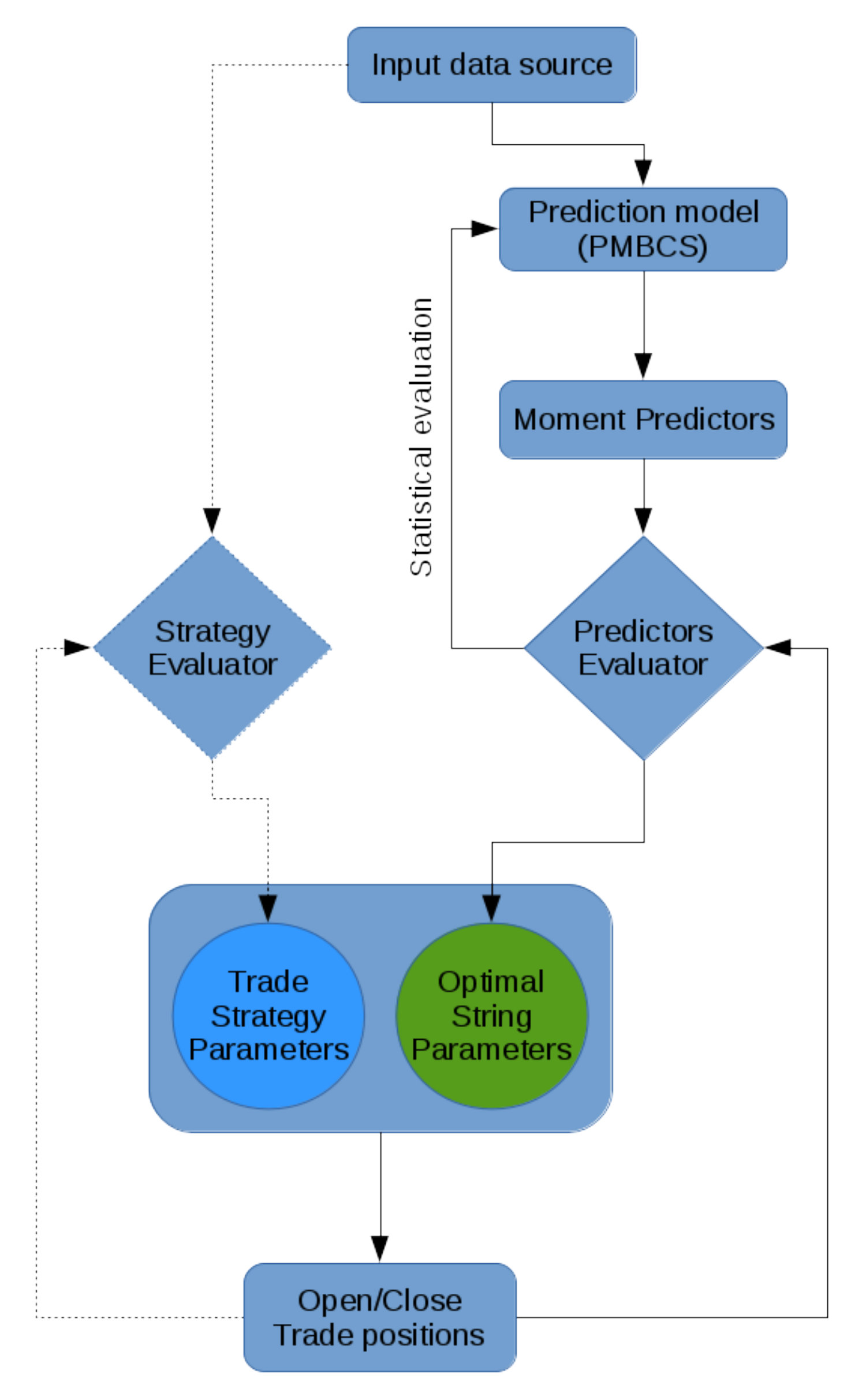}
\end{center}
\caption{The scheme of the trading strategy for the prediction model based on the deviations from the closed string/pattern form.\label{fig:schema}}
\end{figure}

For the PMBCS prediction model the trade system has worked with defined trading strategy, its trading algorithm is schematically reviewed in Fig.~\ref{fig:schema}.

\begin{figure}[!tb]
\begin{center}
\subfloat[]{
	\includegraphics[width=0.48\columnwidth]{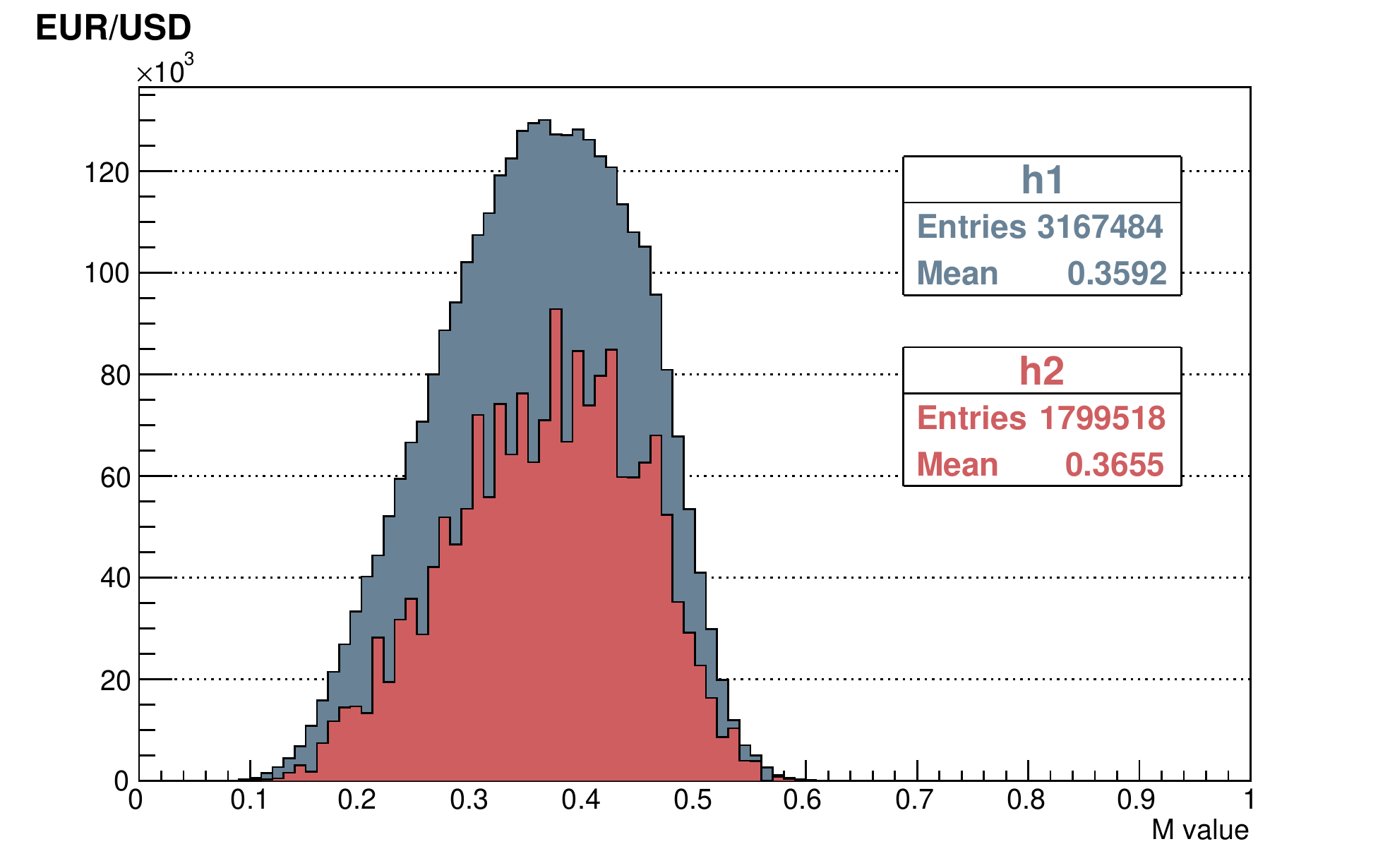}
	\label{fig:histmC}
}
\subfloat[]{
	\includegraphics[width=0.48\columnwidth]{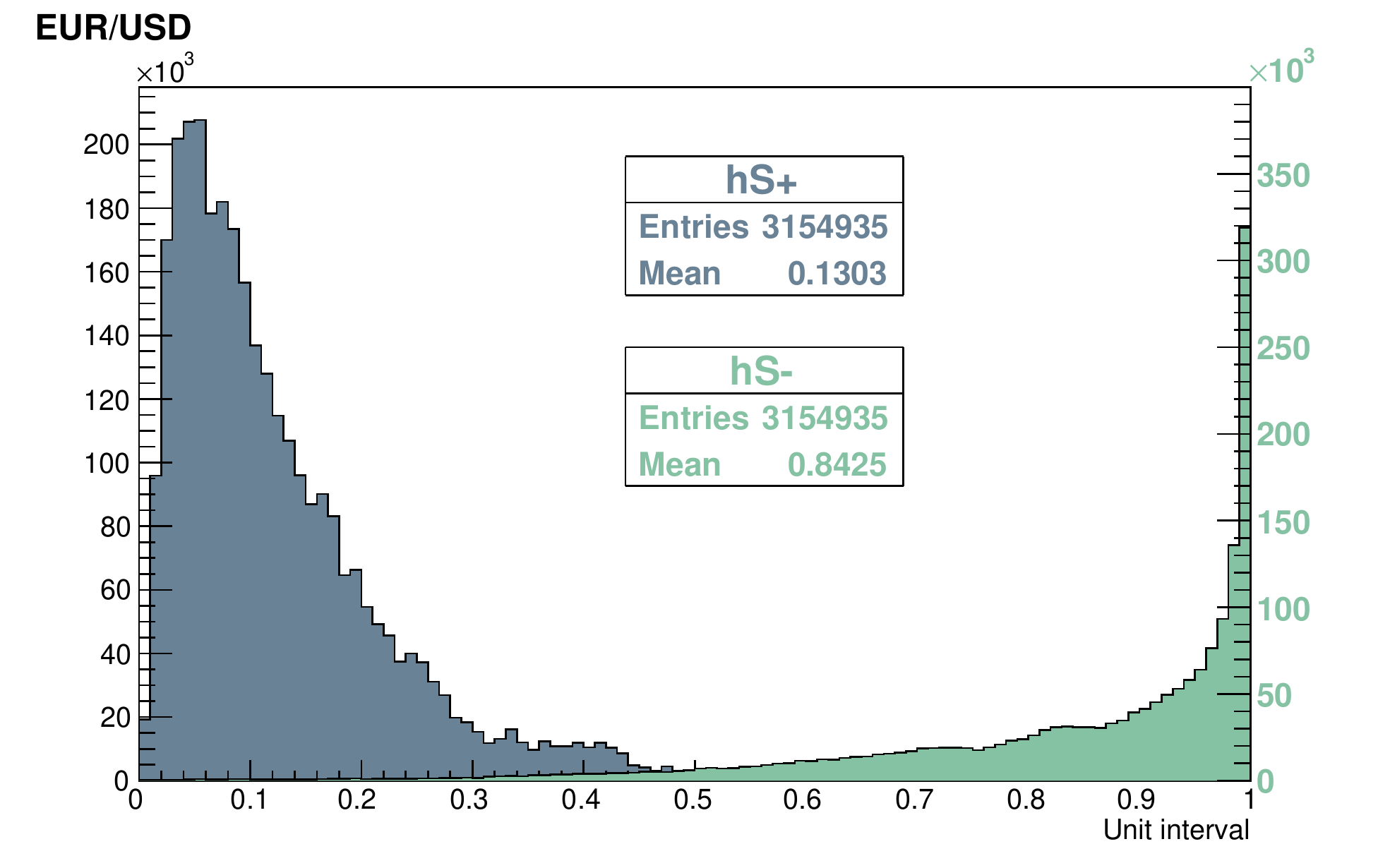}
	\label{fig:histmD}
}
\end{center}
\caption{The typical distributions evaluated by the Predictors Evaluator module. (a) The distributions of the incoming ($h_1$) and outgoing ($h_2$) momenta values $M$. (b) The distributions of $S^{(n)}=+1$ ($h_{S+}$) and $S^{(n)}=-1$ ($h_{S-}$) as dependence on the interval $( h_{\rm cl} - h_{\rm op} )$, both distributions are normalized to the interval $(0,1)$ (for more details see Sec.~\ref{sec:spin}). \label{fig:histS}}
\end{figure}

Various sources serve as input data for the algorithm, e. g., real time data or historical data of currency rates, we have chosen the second one. Data are handled by the current prediction model (PMBCS Prediction Model). In the heart of the algorithm there lies a momentum calculator module (Moment Predictors). The module calculates the values of the momentum predictors, which determine the values for the string parameters of the algorithm (Optimal String Parameters). These values directly affect the trade positions (Open/Close Trade positions). This is the most direct way. But the algorithm works also in the parallel way. The values of the momentum predictors are statistically evaluated with the results of the open/close trade positions and compared with predicted values (Predictors Evaluator). Then the ``optimal'' values of the momenta are sent ahead to provide the ``optimal'' values of the string parameters (Fig.~\ref{fig:histmC}).

Beside the first set of parameters, there exists also the second set of parameters, which controls the risk of the algorithm. They are called the trade strategy parameters and in our case they are represented by a maximum number of simultaneously opened trades, skewness of momenta distribution and Sharpe ratio of closed trades. Together, the trade strategy parameters and ``optimal'' string parameters determine the final opening and closing of trade positions.

The left side of the scheme outlines future Strategy Evaluator. Its purpose is to evaluate the trade strategy parameters and this way to control the risk in a more sophisticated way. However, in this paper, the values of the trade strategy parameters were fixed to reasonable constants throughout the simulations, e.~g., the maximum number of opened trades was set to 10/hour. The main reason is that the online process of finding the best strategy parameters is time consuming and needs intense computing power, which was not able at the time of article preparation. On the other hand, as the trade strategy parameters are the matter of real online trade system, they are not so important at the level of algorithm development.

\begin{figure}[!tb]
\begin{center}
	\includegraphics[width=0.9\columnwidth]{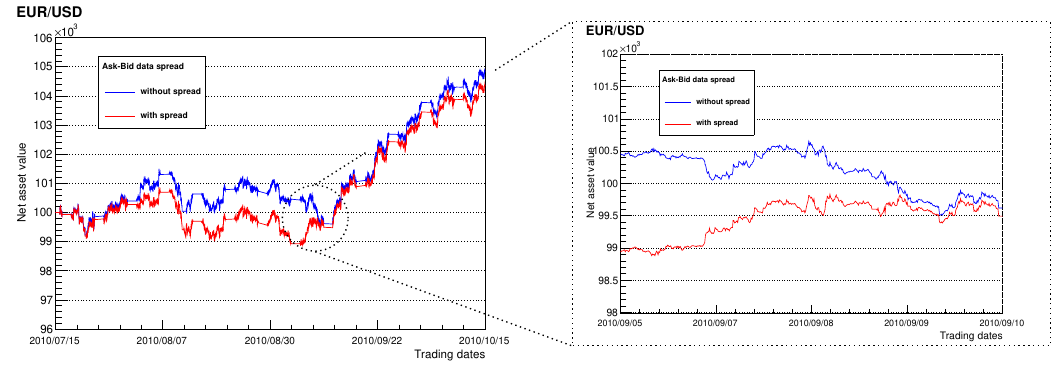}
\end{center}
\caption{The net asset value of the simulation ($l_s=1000$, $Q=1$, $m=1$) on the EUR/USD currency rate for a selected time period as the dependence on the ask-bid data spread.	\label{fig:spread}}
\end{figure}

\begin{figure}[!tb]
\begin{center}
\subfloat[]{
	\includegraphics[width=0.48\columnwidth]{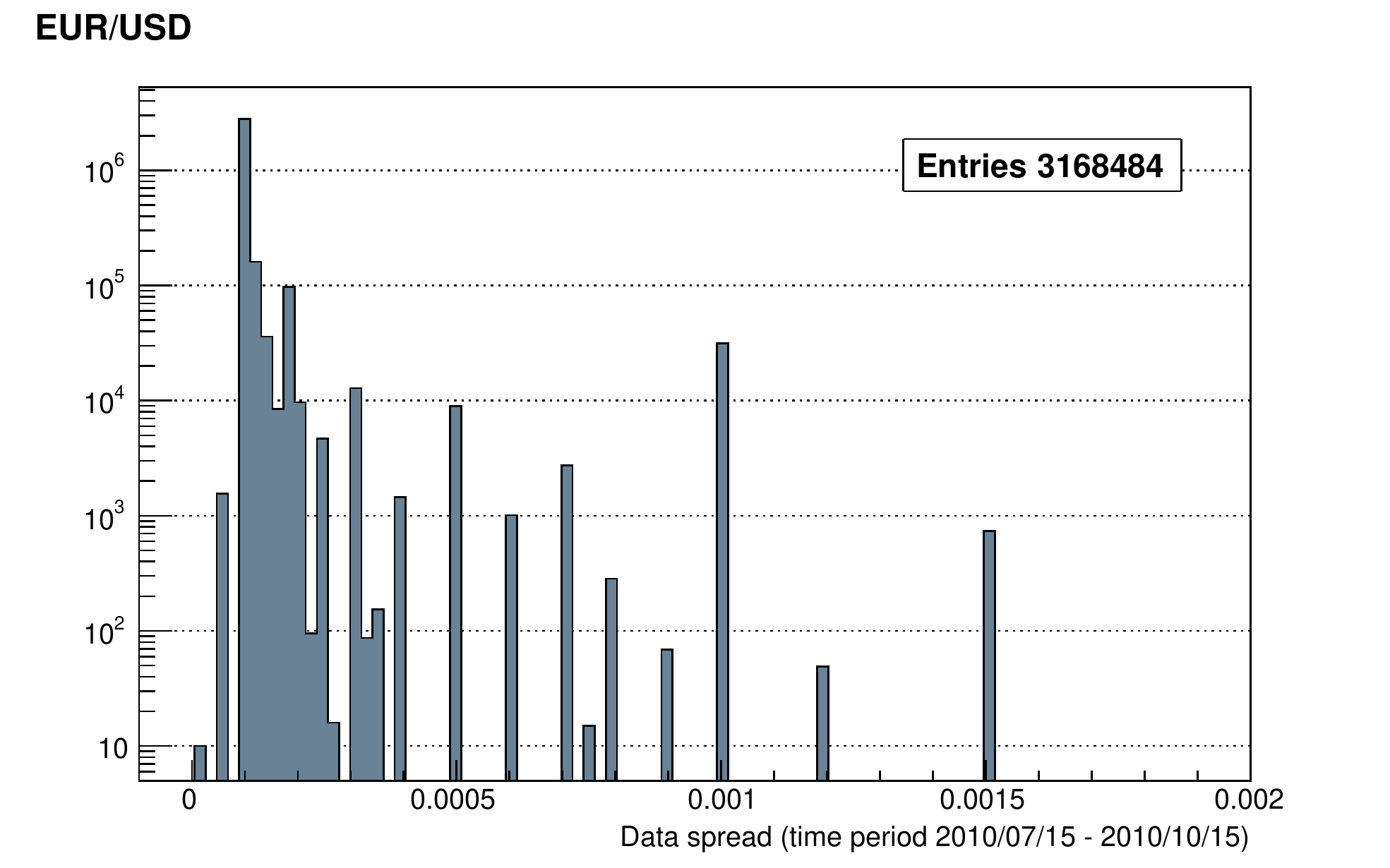}
	\label{fig:histmA}
}
\subfloat[]{
	\includegraphics[width=0.48\columnwidth]{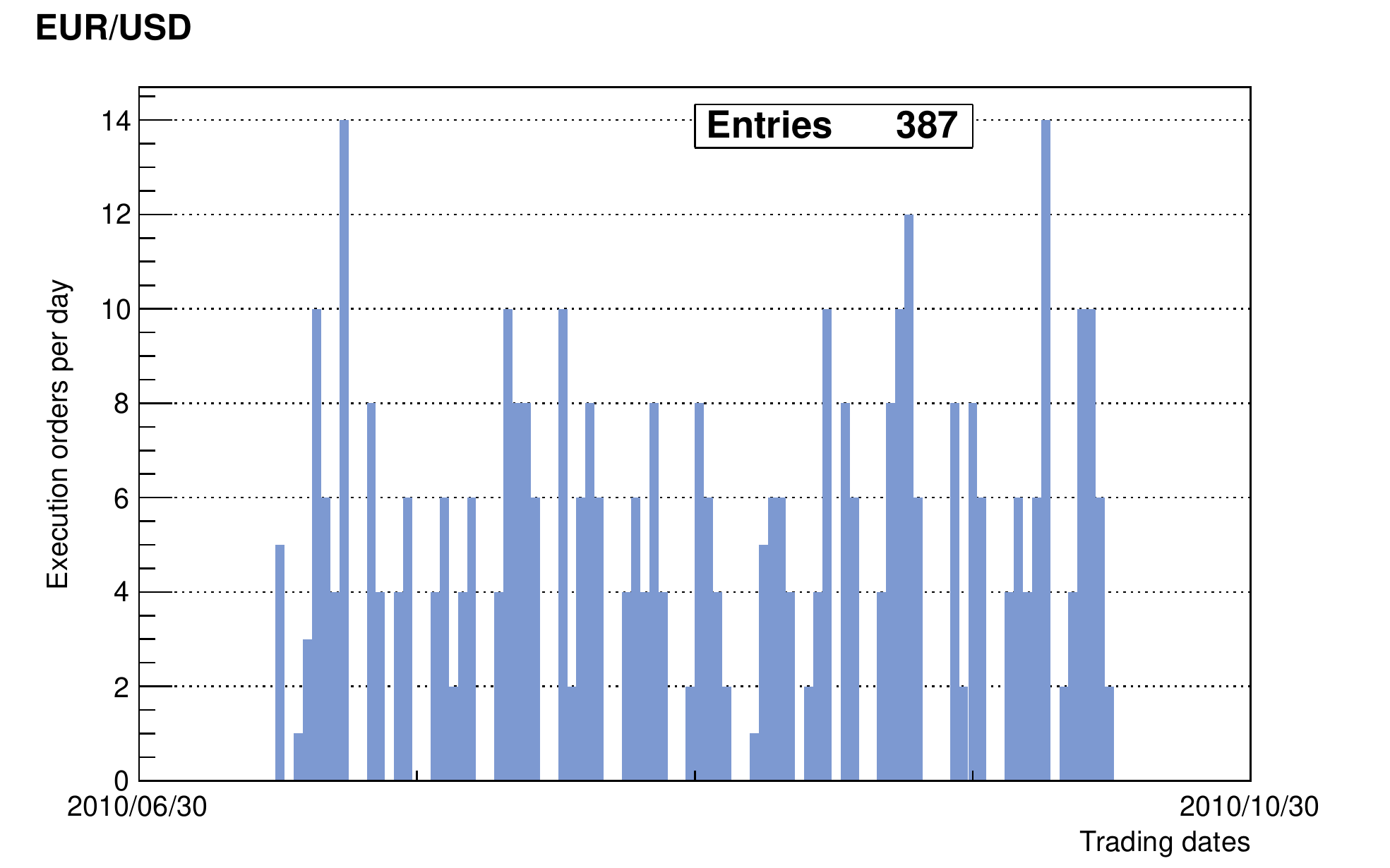}
	\label{fig:histmB}
}
\end{center}
\caption{(a) The ask--bid spread in the evaluated OANDA data, (b) the typical distribution of the execution orders per day. Both histograms are for the time period from 2010/07/15 to 2010/10/15. \label{fig:hist}}
\end{figure}

\section{PMBCS model results}\label{sec:model}

As was mentioned in the previous section, the online trade system operates with two kinds of parameters, there are trade strategy parameters and optimal string parameters. We have not pretended to determine all parameters, as the process of the online trading is robust and it requires a verified and confident system. Our purpose was to demonstrate how the string parameters influence the results and to find out the method how they can be optimized in a simulation process. However, the trade strategy parameters were chosen as close as possible to the real ones.

\subsection{Empirical analyses} \label{subsec:analyses}

As the starting point of our prediction simulations we considered the string parameters which describe the momentum function $M_{(l_s, m, Q, F_{\m{CS}})}$ (see Eq~(\ref{eq:momentum})): $l_s$ -- the length of momentum string (the number of ticks or time period), $Q$ -- the quotient or the exponent of the momentum (a deformation of string), $m$ -- the frequency of the momentum function and last but not least, the momentum function depends also on the $F_{\m{CS}}(h,l_s)$ -- the regular function. We illustrated the impact of these parameters and the prediction behavior of our model on the net asset value.

The simulations were carried out on the OANDA data for EUR/USD currency rate for the time period of three months, from 2010/07/15 to 2010/10/15. The typical background of the simulations is presented in Figs.~\ref{fig:histS}--\ref{fig:hist}. Figure~\ref{fig:spread} shows the dependence in the three month simulation on the ask-bid data spread, i.~e., on the transaction costs (Fig.~\ref{fig:histmA}). Also for long trade periods our algorithm behaves very stably, the values of NAV do not show the high dependence of the transaction costs. The distribution of the trades is nearly uniform throughout all time period, it varied in the range from $0$ to $14$ per day (Fig.~\ref{fig:histmB}). There are no rapid increases and decreases of the amounts of the trades.

The string statistics in the Predictors Evaluator module is represented by Fig.~(\ref{fig:histmC}) by the typical distribution of evaluated momenta $M$. The distributed momenta incoming and outgoing from the module are represented by the histograms $h_1$ and $h_2$. The figure shows that the selected momenta $M$ with the expected values (the mean) from the interval $(0.3, 0.4)$ are actually crucial in the predictions. Their treatment in the first steps of the evaluation can define the next course of the simulation and to shorten the computation time.

\begin{figure}[!tb]
\begin{center}
\subfloat[]{
	\includegraphics[width=0.48\columnwidth]{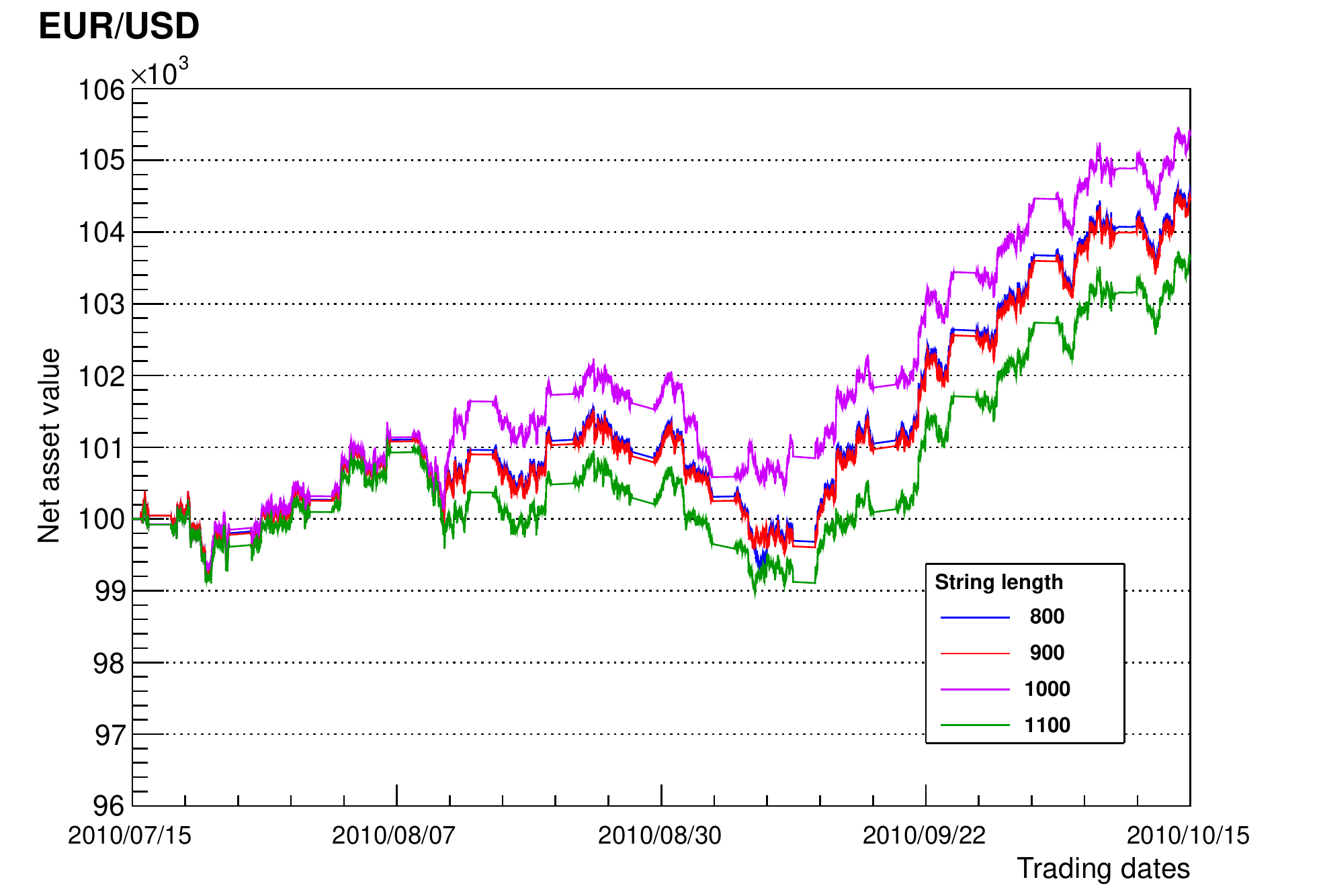}
	\label{fig:ls}
}
\subfloat[]{
	\includegraphics[width=0.48\columnwidth]{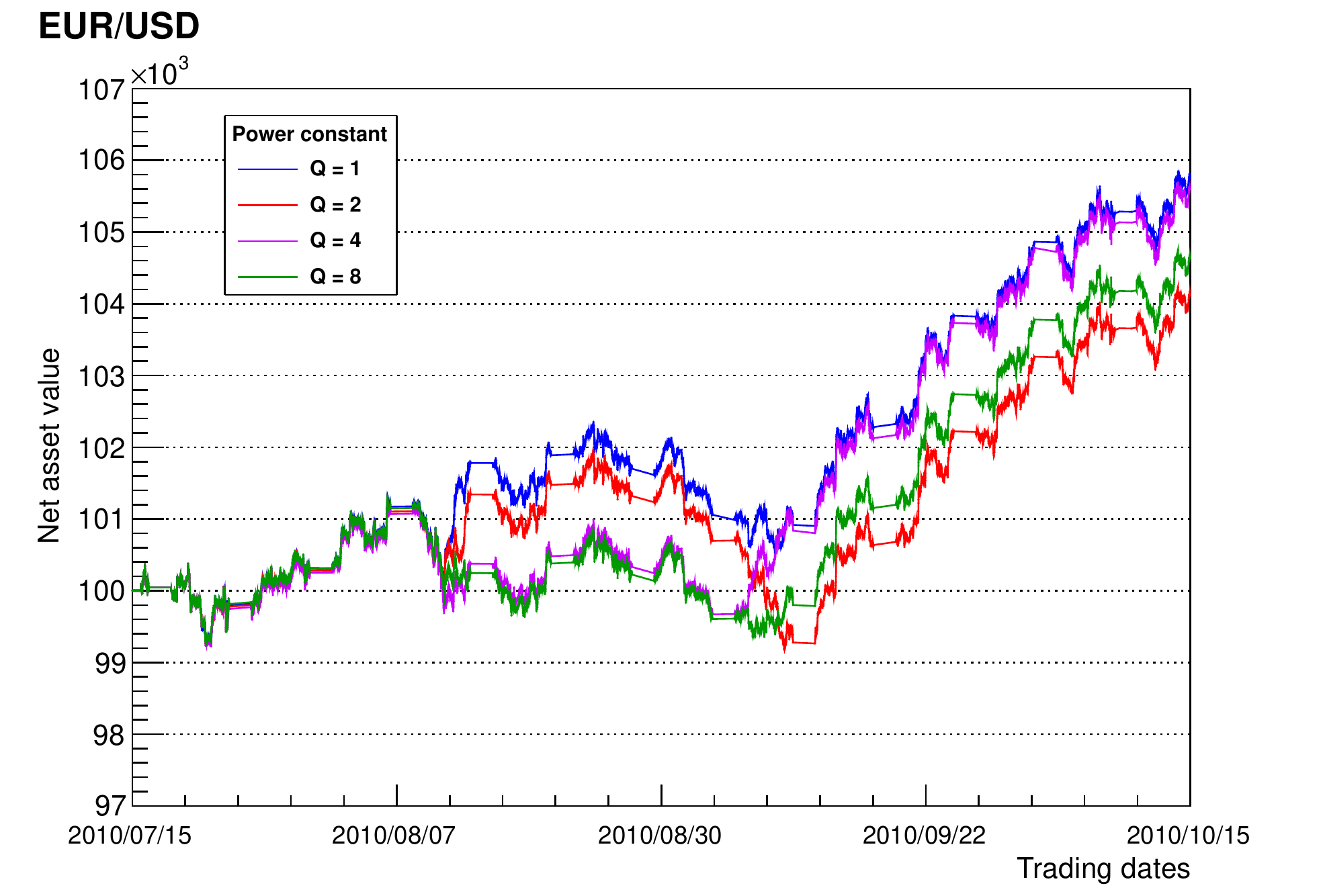}
	\label{fig:powerA}
}
\end{center}
\caption{The net asset value of the model on the EUR/USD currency rate for a selected time period as the dependence on the string length parameter and the power constant.\label{fig:par}}
\end{figure}

\begin{figure}[!htb]
\begin{center}
\subfloat[]{
	\includegraphics[width=0.48\columnwidth]{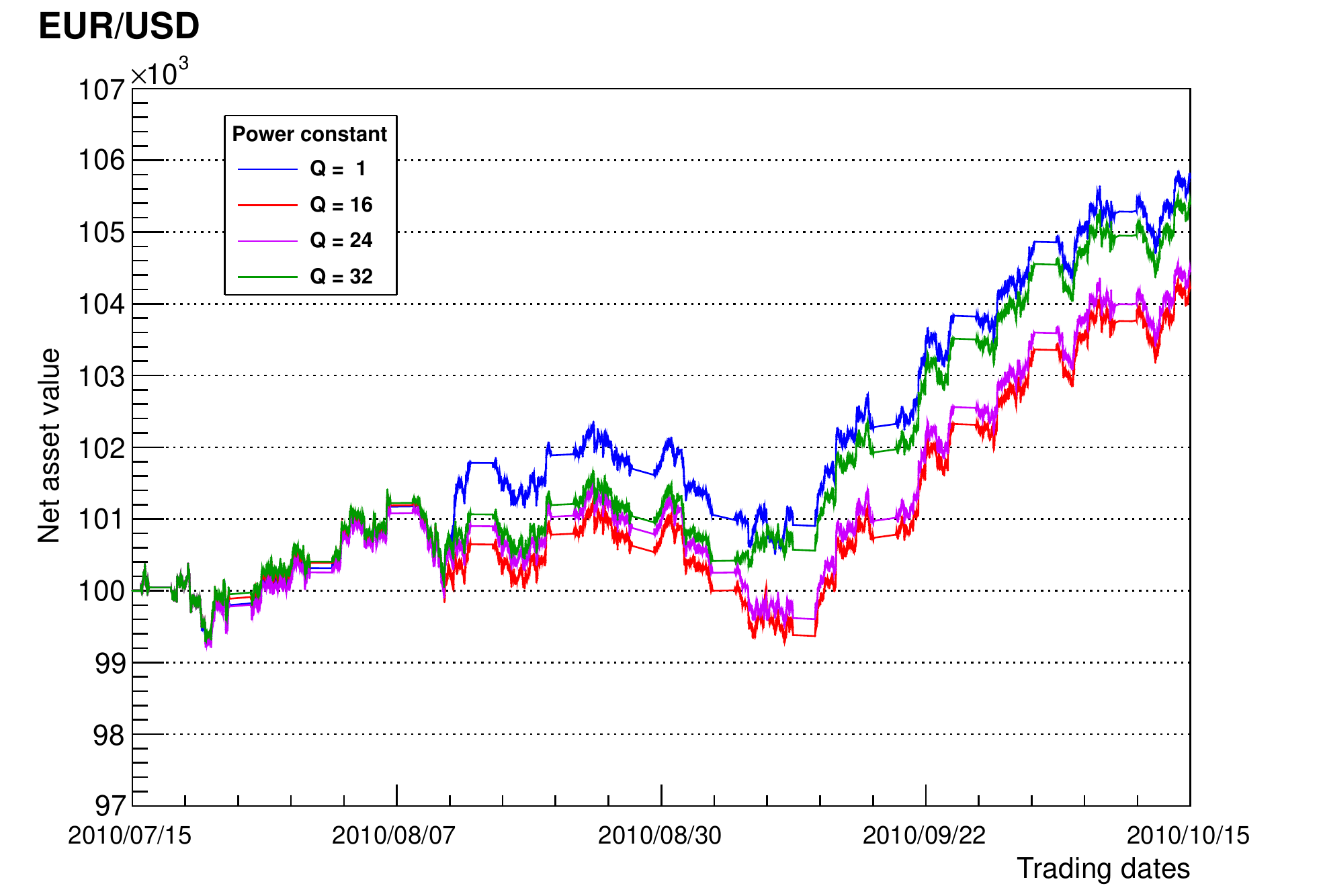}
	\label{fig:powerB}
}
\subfloat[]{
	\includegraphics[width=0.48\columnwidth]{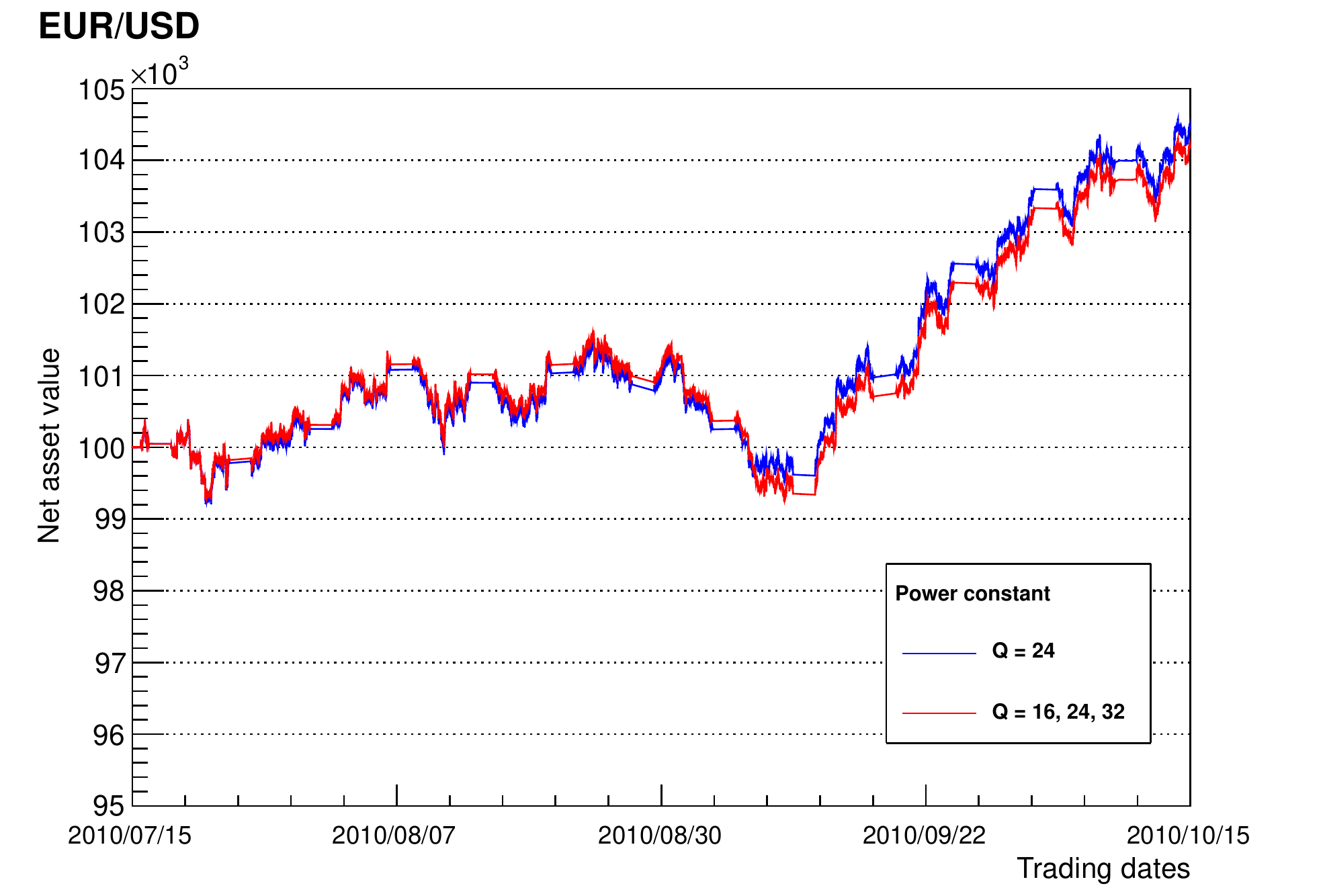}
	\label{fig:powerC}
}
\end{center}
\caption{The net asset value of the model on the EUR/USD currency rate for a selected time period as the dependence on the power constant.}
\end{figure}

\begin{figure}[!tb]
\begin{center}
\subfloat[]{
	\includegraphics[width=0.48\columnwidth]{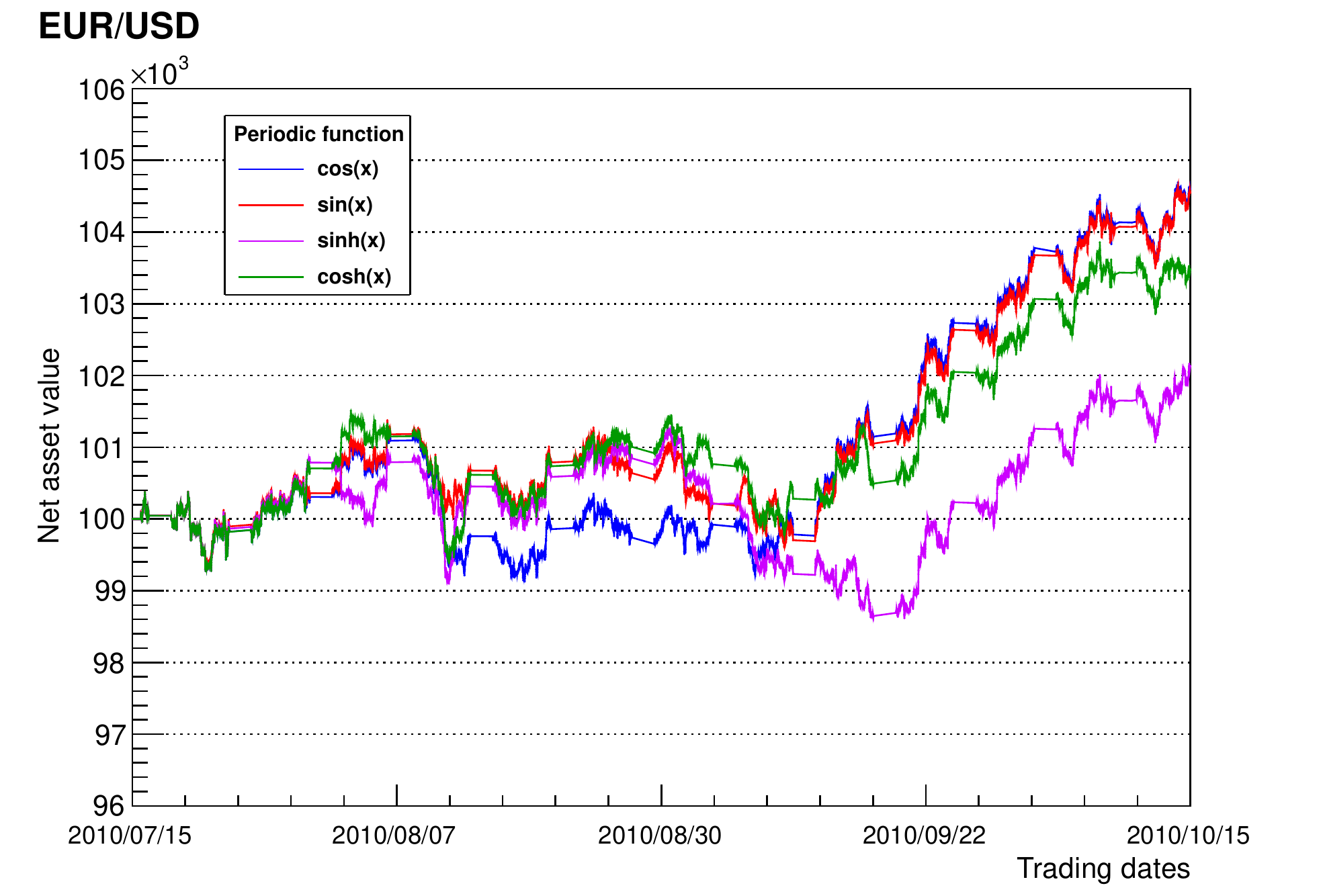}
	\label{fig:funcAlla}
}
\subfloat[]{
	\includegraphics[width=0.48\columnwidth]{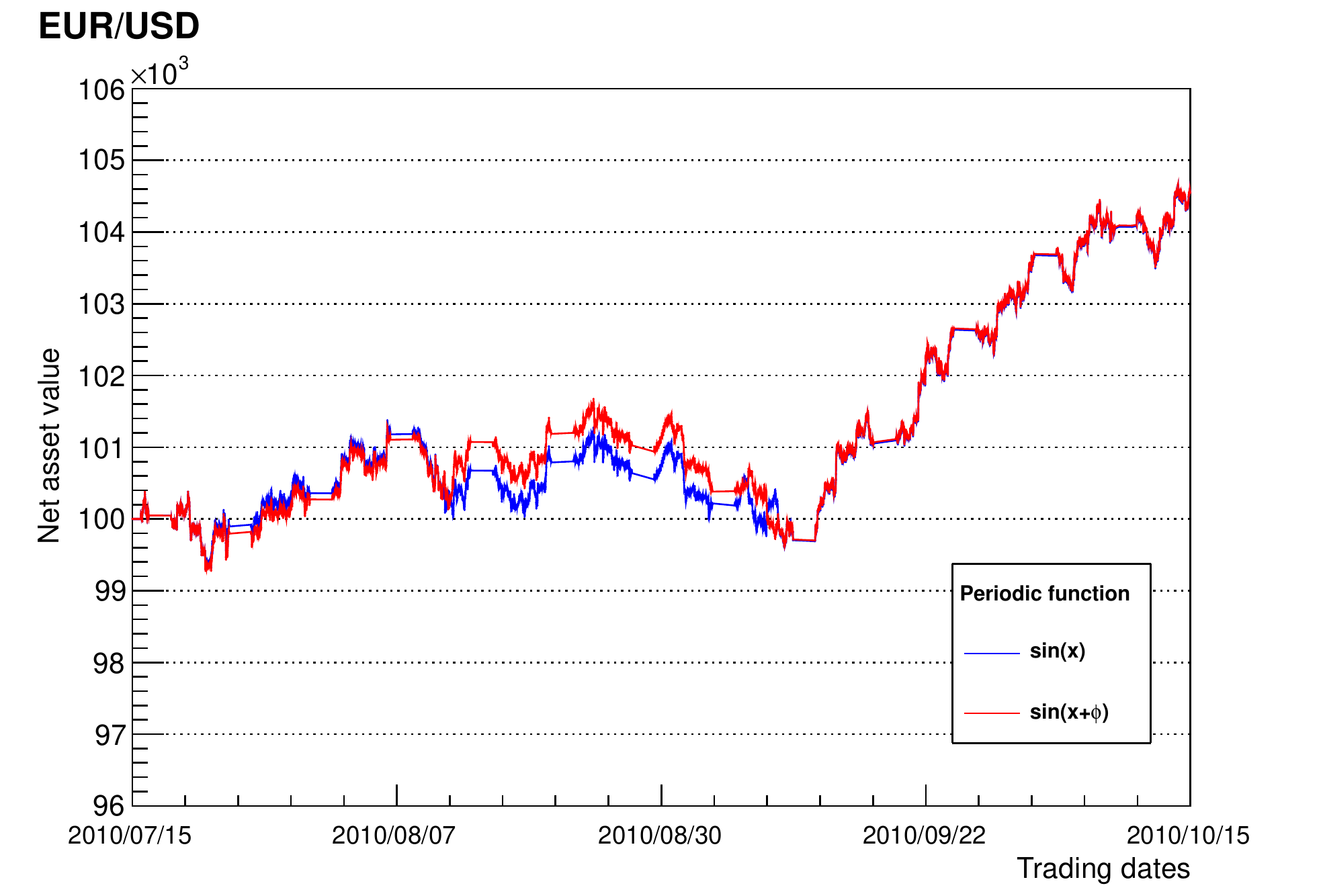}
	\label{fig:funcAllb}
}
\end{center}
\caption{The net asset value of the model on the EUR/USD currency rate for a selected time period as the dependence on the periodic function $F_{\m{CS}}$, (a) the comparison of the forecasts for the functions $\cos(x)$, $\sin(x)$, $\sinh(x)$, $\cosh(x)$, (b)  the comparison of the forecasts for the functions $\sin(x+\phi)$, where $\phi=0,\pi$.}\label{fig:funcAll}
\end{figure}

In Figs.~\ref{fig:par}--\ref{fig:funcAll} we described some interesting results relating to the string parameters.
The net asset value of the model dependence on the string length parameter $l_s$ is presented in Fig.~\ref{fig:ls}. As one can see, the value of $l_s=900$ seems to be most promising, this value was fixed for the next predictions.
The dependence of the model on the parameter $Q$ -- the quotient of the moment is described on the next figures. Figure~\ref{fig:powerA} represents the dependence on low values, i.~e., $Q=1,2,4,8$, Fig.~\ref{fig:powerB} represents the dependence on higher values, i.~e., $Q=1,16,24,32$. The comparison of the value $Q=24$, which seems to be most suitable for the next forecasts, with the simultaneous use of three values is shown in Fig.~\ref{fig:powerC}.

The interesting case is the choice of the regular function $F_{\m{CS}}$. The previous forecasts \cite{P:1,P:2} were made for a trigonometric function $\cos(x)$. In this paper, we presented the results of the tests with other functions, as one can see in Fig.~\ref{fig:funcAll}. The comparisons of the forecast with the function $\cos(x)$ were made for forecasts with functions $\sin(x)$, $\sinh(x)$ and $\cosh(x)$ in Fig.~\ref{fig:funcAlla}. The subfigure Fig.~\ref{fig:funcAllb} represents the comparison of forecasts for the function $\sin(x)$ with the different arguments, i.~e., $x$ and $x+\phi$, where $\phi=0,\pi$.

\subsection{Self-learning model}\label{subsec:self}

The presented forecasts only partially show the possibilities of our algorithm. The values of string parameters, which were used in previous forecasts, are summarized in Table~\ref{tab:string} in a simple model column. For each forecast we used one value for each type of parameters: $l_s$, $Q$, $F_{\m{CS}}$, $m$ and $\phi$, respectively. In other words, only one set of string parameters is used, we have denoted it by $n_s=1$. However, the algorithm can work with various sets of free parameters simultaneously ($n_s\ge1$). This is possible due to the fact that the model was enhanced to the so called Self-learning model. In terms of the string parameters from Table~\ref{tab:string}, all possible combinations of the values from the third column are taken into account and the corresponding momentum predictors are calculated.

\begin{table}
\begin{tabular}{@{}ccc}
\toprule
\textbf{String parameters} & \textbf{PMBCS Simple model} & \textbf{PMBCS Self-learning model}\\
\midrule
$l_s$ & 800, 900, 1000, 1100 & [900] \\
$Q$ & 1, 2, 4, 8, 16, 24, 32 & [8, 16, 24, 32]  \\
$F_{\m{CS}}$  & $\cos(x)$, $\sin(x)$, $\sin(x+\phi)$, $\sinh(x)$, $\cosh(x)$ & $[\cos(x+\phi)]$ \\
$m$ & 0, 1, 2, 3 & [0, 1, 2, 3] \\
$\phi$ & $0$, $3.14$ & [$0$, $3.14$] \\
\bottomrule
\end{tabular}
\caption{The values of string parameters used for the PMBCS Simple and Self-learning models. The square brackets emphasize the fact that the Simple model works with exactly one value of the analyzed string parameter values, while the Self-learning model can work with sets of parameters simultaneously.\label{tab:string}}
\end{table}

As was said before in Sec.~\ref{subsec:strategy}, the values of momentum predictors under consideration are statistically evaluated (Predictor Evaluator module). For this purpose, the statistical quantity called the Sharpe ratio was introduced \cite{P:3}, it is defined as
\begin{gather} \label{eq:sr}
S=\frac{E(R-R_f)}{\sigma},
\end{gather}
where $R$ is the asset return, $R_f$ is the return on a benchmark asset (risk free), $E(R-R_f)$ is the mean value of the excess of the asset return over the benchmark return, $\sigma$ is the standard deviation of the excess of the asset return. The definitions of formulas are
\begin{align}
E(R-R_f)=&\frac{1}{N}\sum\limits_{i=1}^{N}(R^{[i]}-{R_f}^{[i]}),\quad R^{[i]}=\sum\limits_{j=1}^{T^{[i]}}p_j,\quad {R_f}^{[i]}=\sum\limits_{j=1}^{T^{[i]}}(p_j-j*P),\\
\sigma=&\sqrt{\frac{1}{N}\sum\limits_{i=1}^{N}{(R^{[i]}-{R_f}^{[i]})}^2},
\end{align}
where $N$ is the number of the total closed positions, $T^{[i]}$ is the number of the accepted trade results (closed positions), $P$ is the unit penalty parameter. The closing of the trade position is influence by the trade altitude parameter. Its value as well the value of the unit penalty parameter $P$ were set empirically.

\begin{figure}[!htb]
\begin{center}
\subfloat[]{
	\includegraphics[width=0.48\columnwidth]{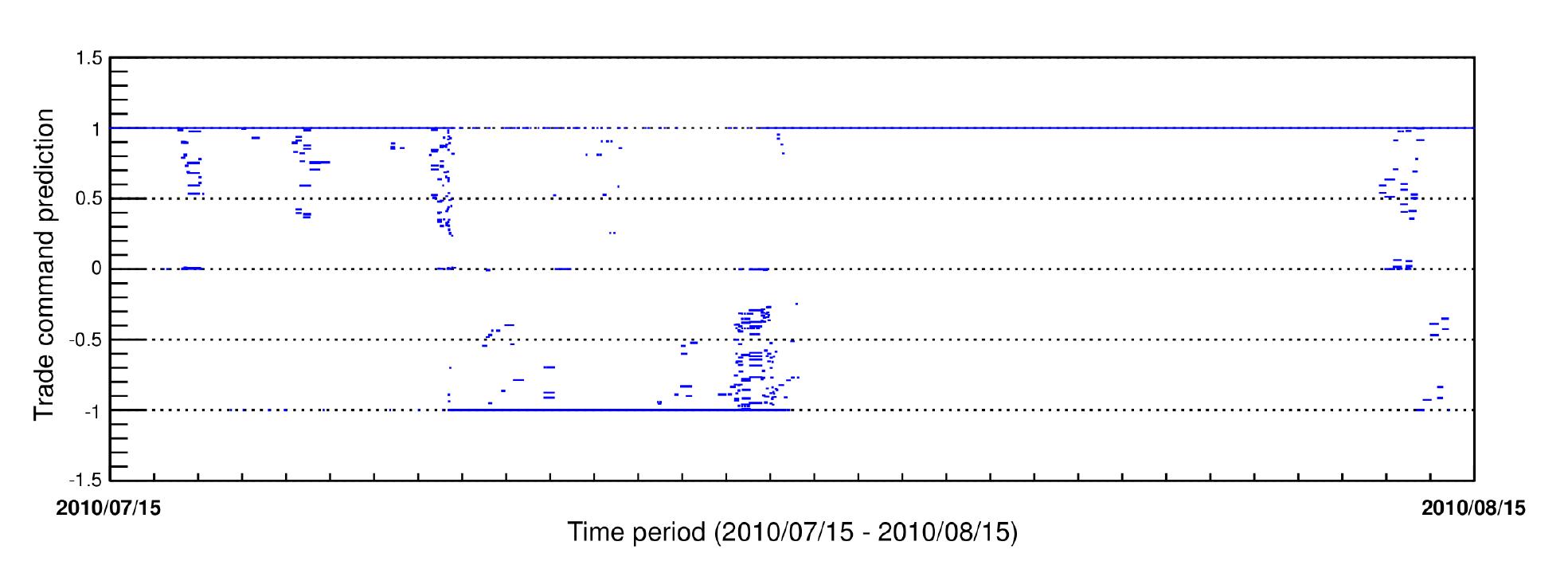}\label{fig:sharpA1}
}
\subfloat[]{
	\includegraphics[width=0.48\columnwidth]{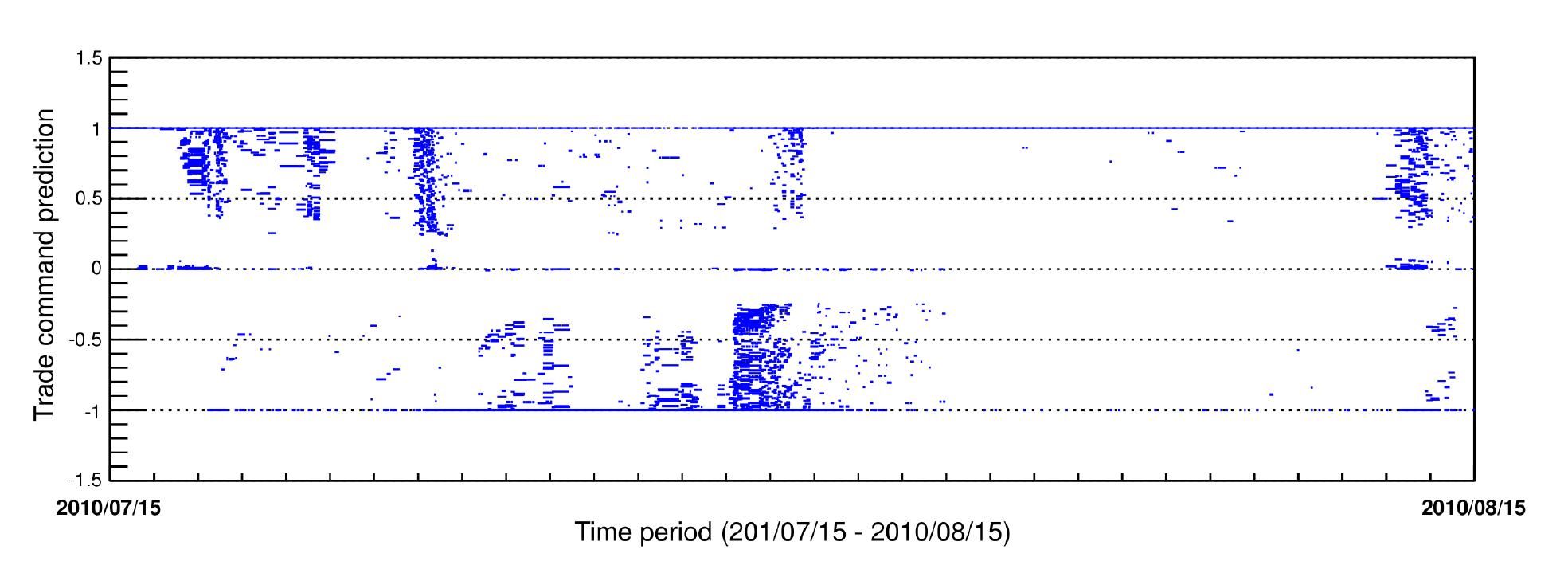}\label{fig:sharpA2}
}\\
\subfloat[]{
	\includegraphics[width=0.48\columnwidth]{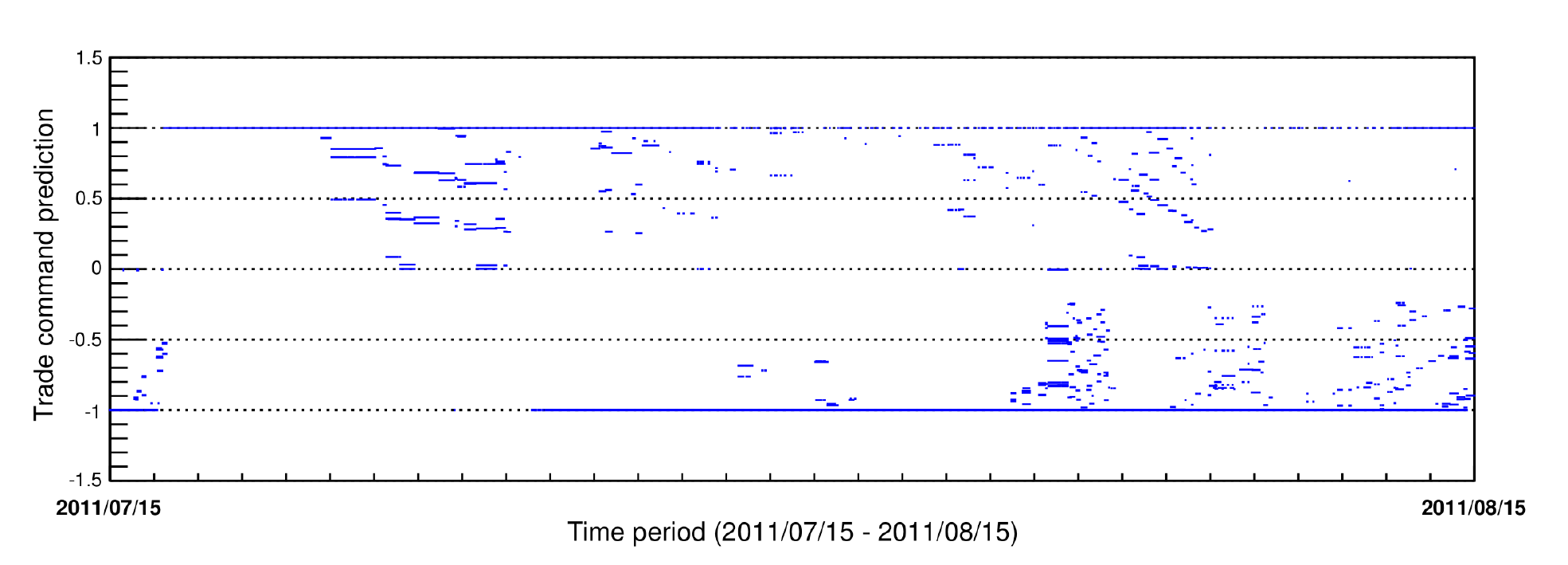}\label{fig:sharpB1}
}
\subfloat[]{
	\includegraphics[width=0.48\columnwidth]{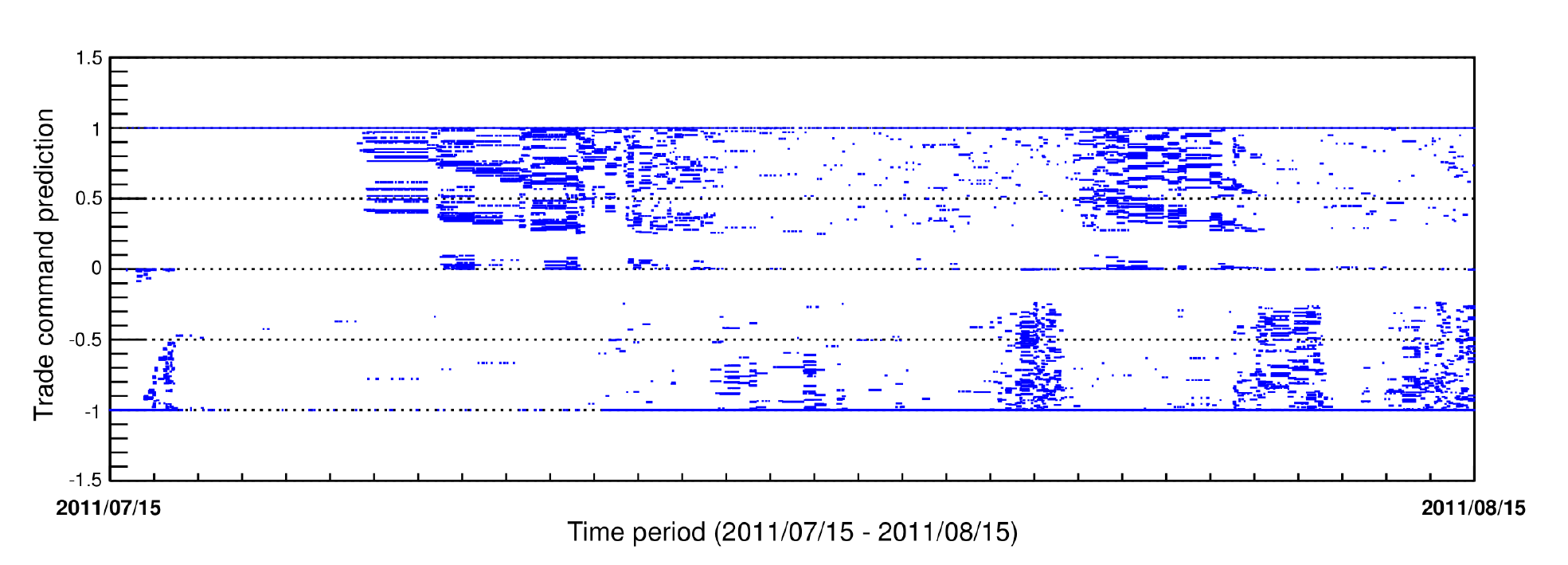}\label{fig:sharpB2}
}
\end{center}
\caption{The split of the trade command predictions for selected time periods as the dependence on the number of parallel predictor computations $n_s$, left column (a),(c) with $n_s=1$, right column (b),(d) with $n_s=8$ .\label{fig:sharp}}
\end{figure}

The Sharpe ratio helps to find the best combinations of string parameters. In a such way, the model is the Self-learning model, the trading works with the optimal parameters up to the moment when the new optimal combination of parameters is found. For this purpose, the Sharpe ratio needs a sufficient amount of predicted momenta to provide reliable optimization. The number of sets of string parameters $n_s$ must be fixed at a sufficient number, as one can see from Fig.~\ref{fig:sharp} where the split of the trade command predictions is demonstrated for $n_s=1$ and $n_s=8$. The positive values of trade command prediction are in favour of the trade, the negative values are against the trade.

\begin{figure}[!tb]
\begin{center}
\subfloat[]{
	\includegraphics[width=0.48\columnwidth]{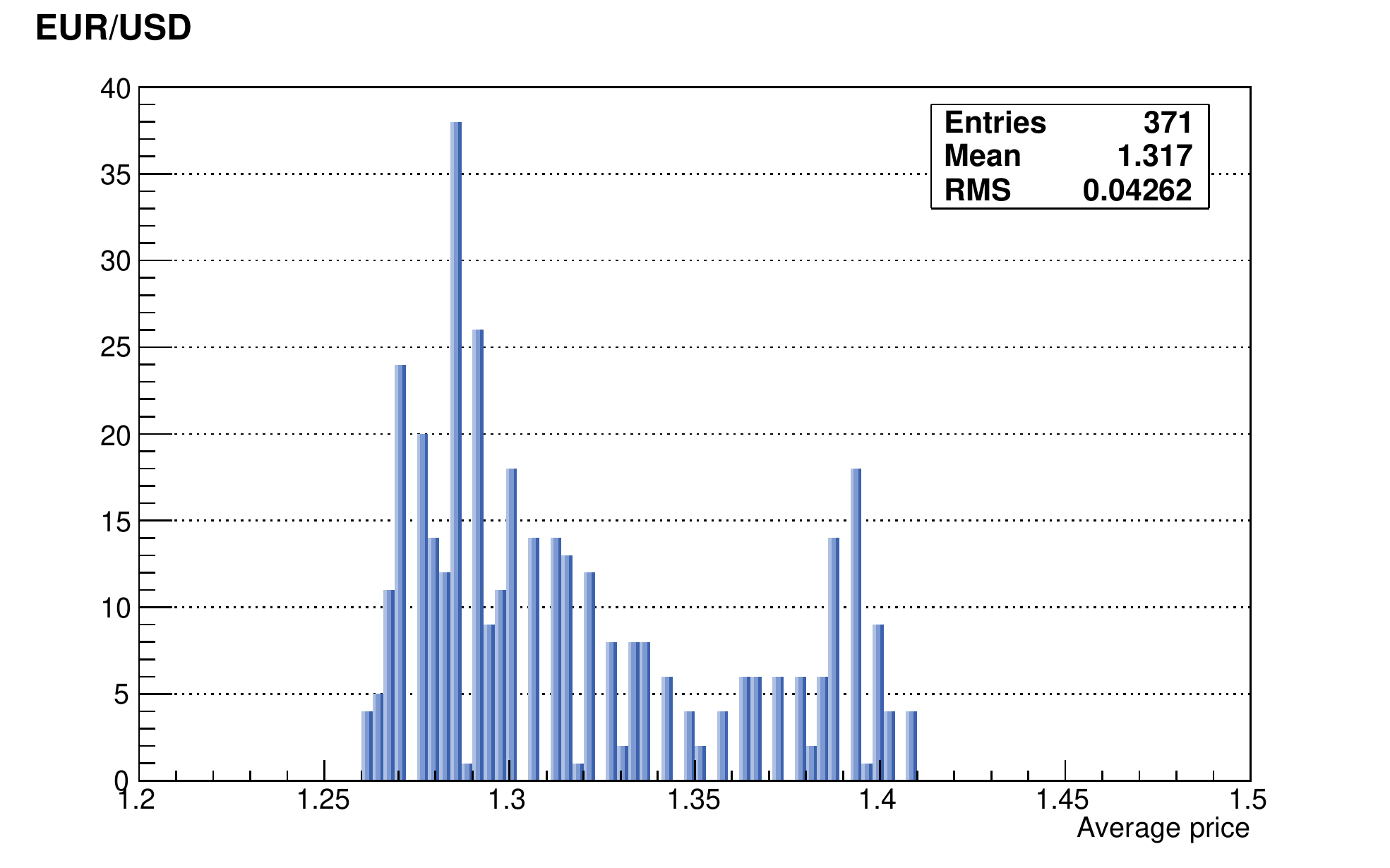}\label{fig:src1}
}
\subfloat[]{
	\includegraphics[width=0.48\columnwidth]{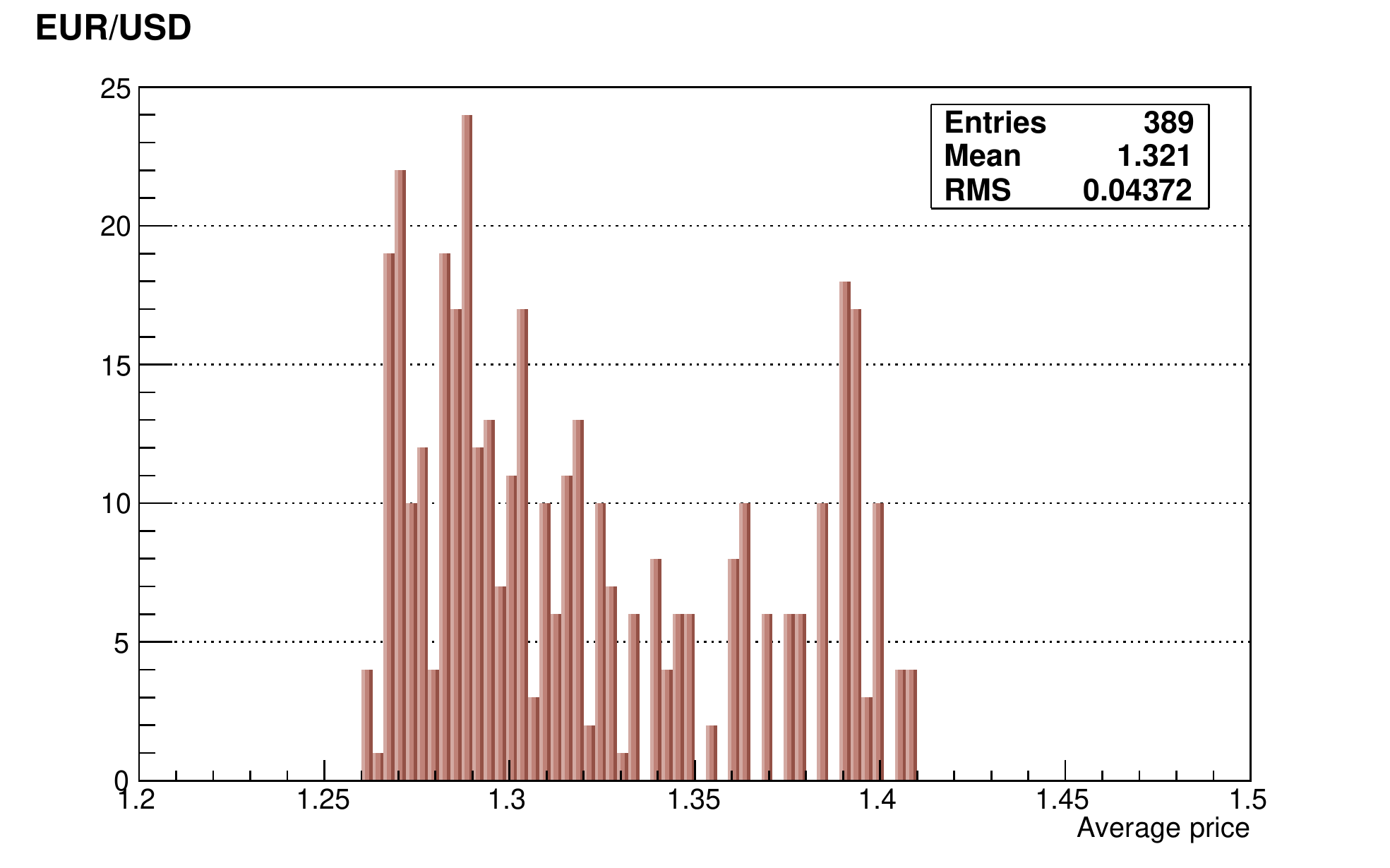}\label{fig:src2}
}\\
\subfloat[]{
	\includegraphics[width=0.48\columnwidth]{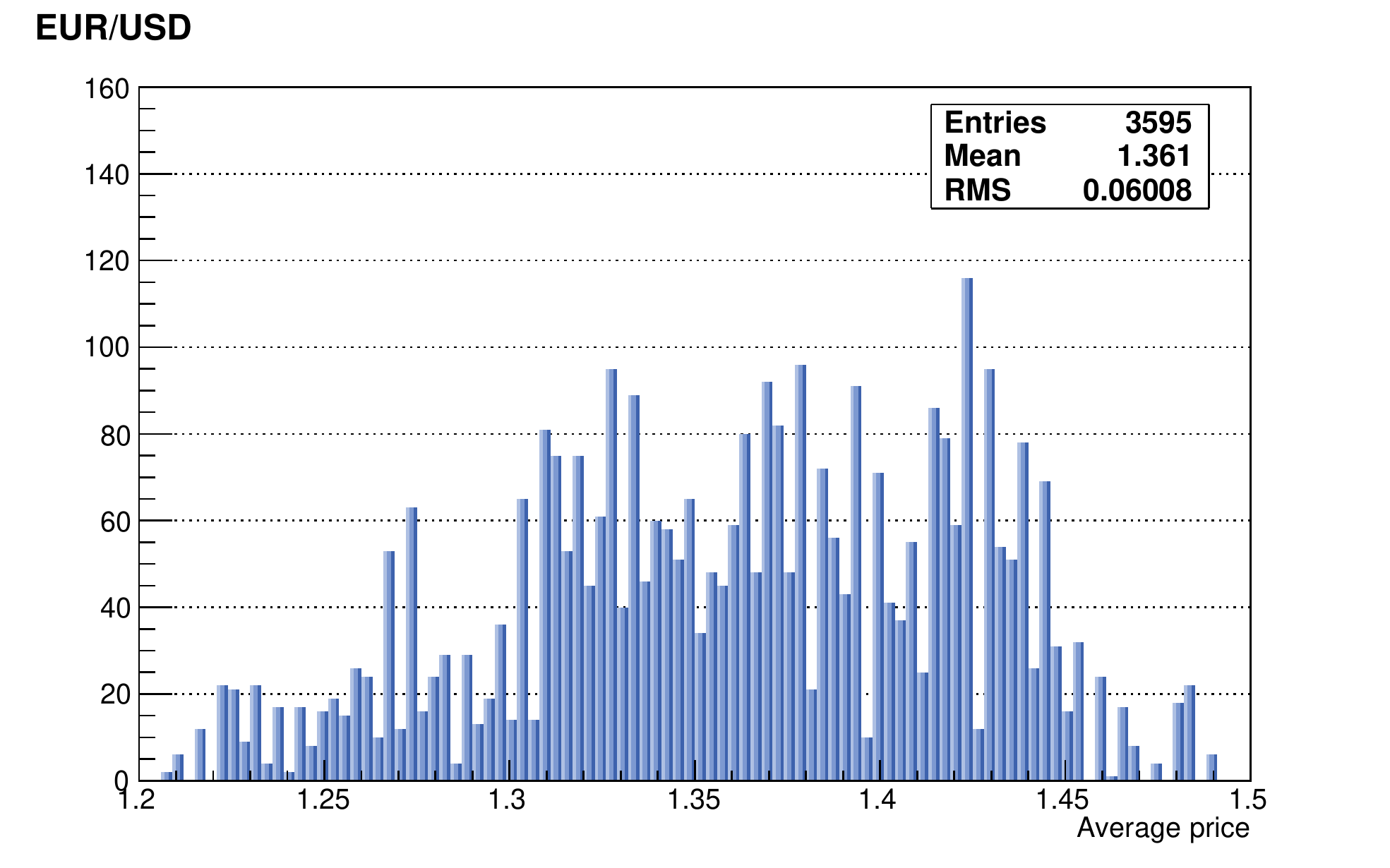}\label{fig:src3}
}
\subfloat[]{
	\includegraphics[width=0.48\columnwidth]{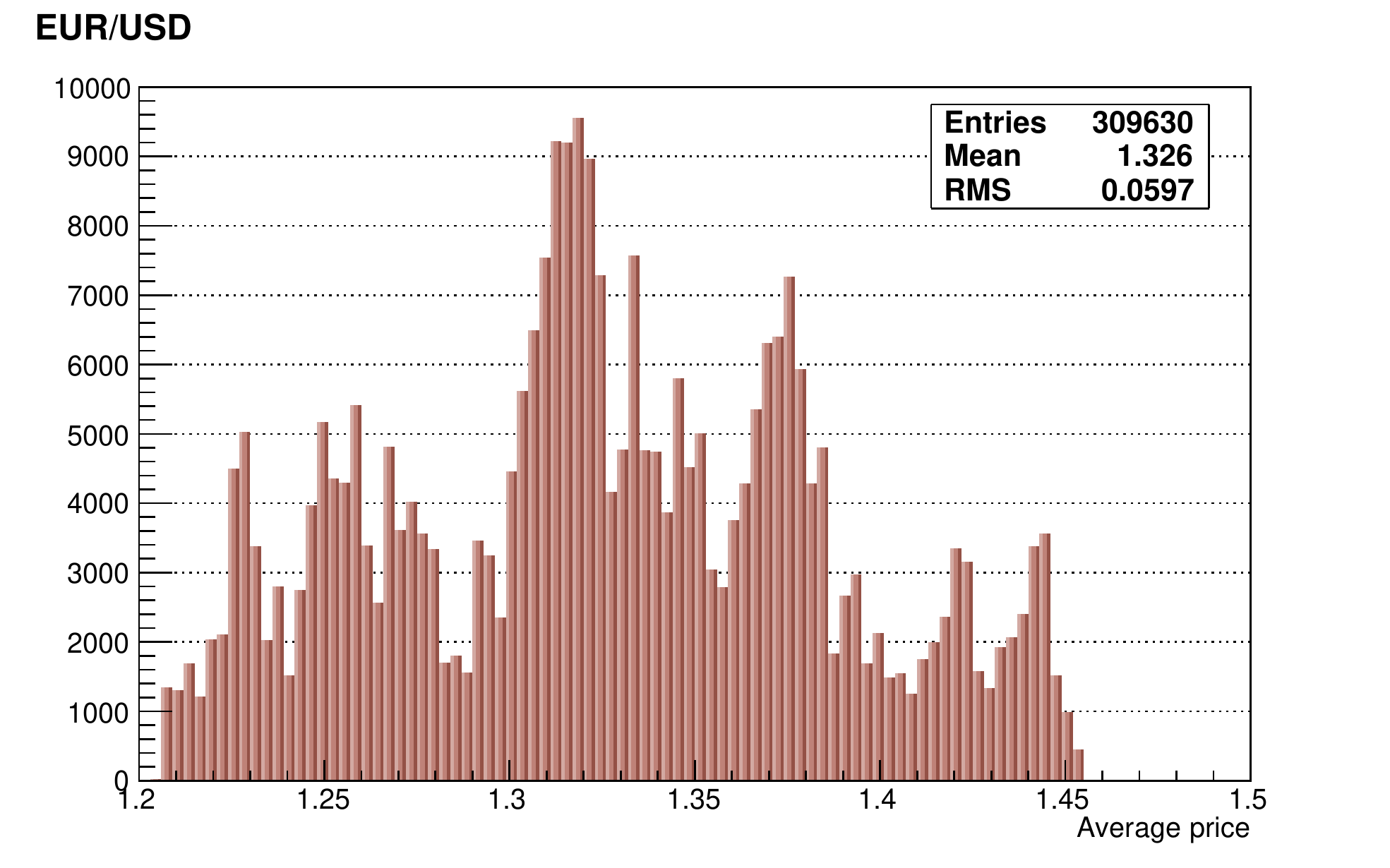}\label{fig:src4}
}

\end{center}
\caption{The histograms of the average values of the execution reports for the Self-learning model with $n_s=2$ (left column) and $n_s=16$ (right column). The time period of the simulation is three months for (a), (b) and one year for (c), (d).\label{fig:src}}
\end{figure}

The lower value of $n_s$ means less statistics available for the finding of optimal parameters. On the other hand, very high values of $n_s$ need the corresponding computing power. In Fig.~\ref{fig:src}, we compared the effect of the higher value of $n_s$ for the Self-learning model with $n_s=2$ (left column) and $n_s=16$ (right column). The histograms show the average values of the execution reports sent by the model. In the subfigures \ref{fig:src1}--\ref{fig:src2} (three months data), the effect is not seen, however, in the subfigures \ref{fig:src3}--\ref{fig:src4} (one year data) the number of the execution reports is approximately 100 times higher. It means that the algorithm is more flexible in trading for a longer period.

\begin{figure}[!htb]
\begin{center}
\subfloat[]{
	\includegraphics[width=0.48\columnwidth]{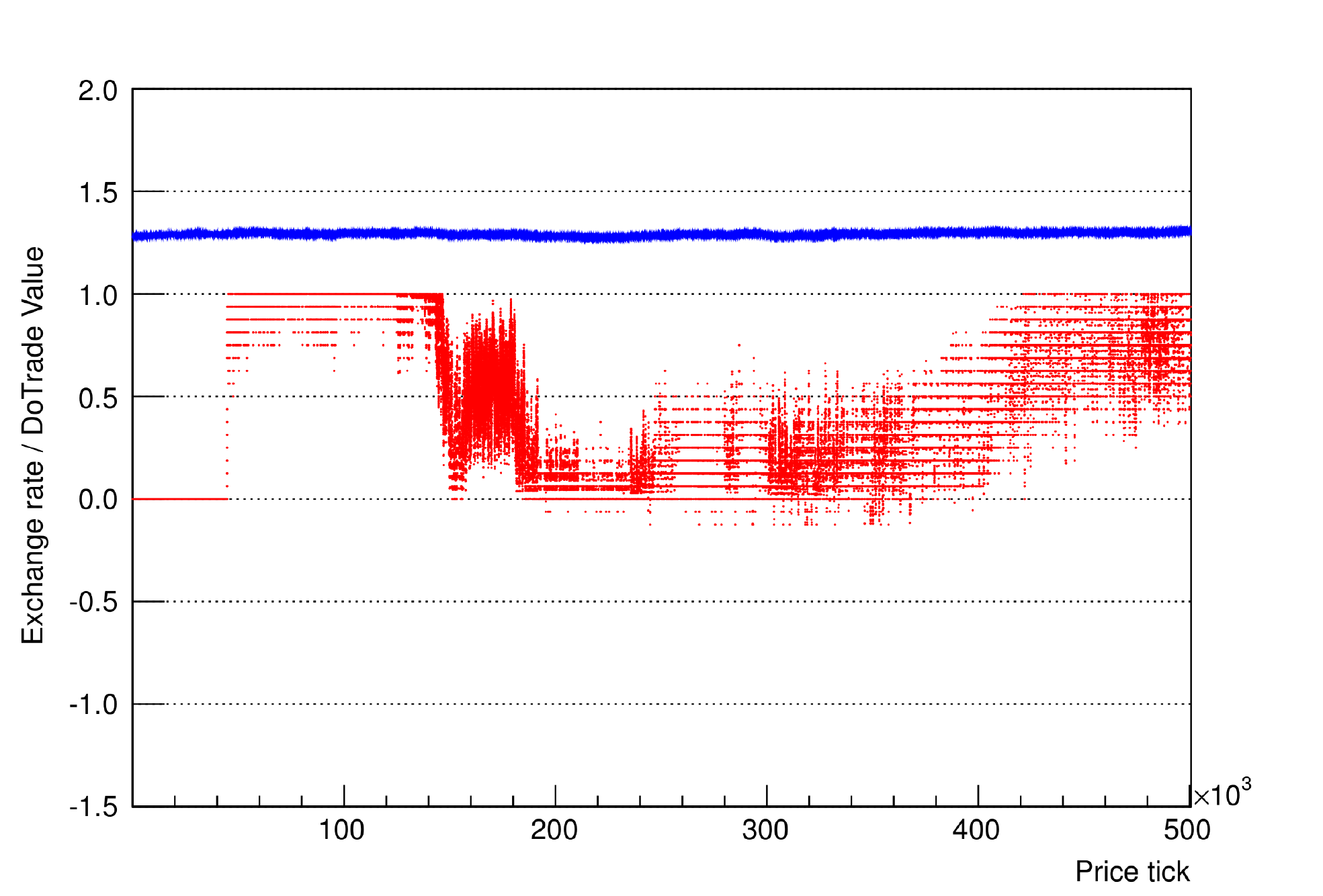}
	\label{fig:PredA}
}
\subfloat[]{
	\includegraphics[width=0.48\columnwidth]{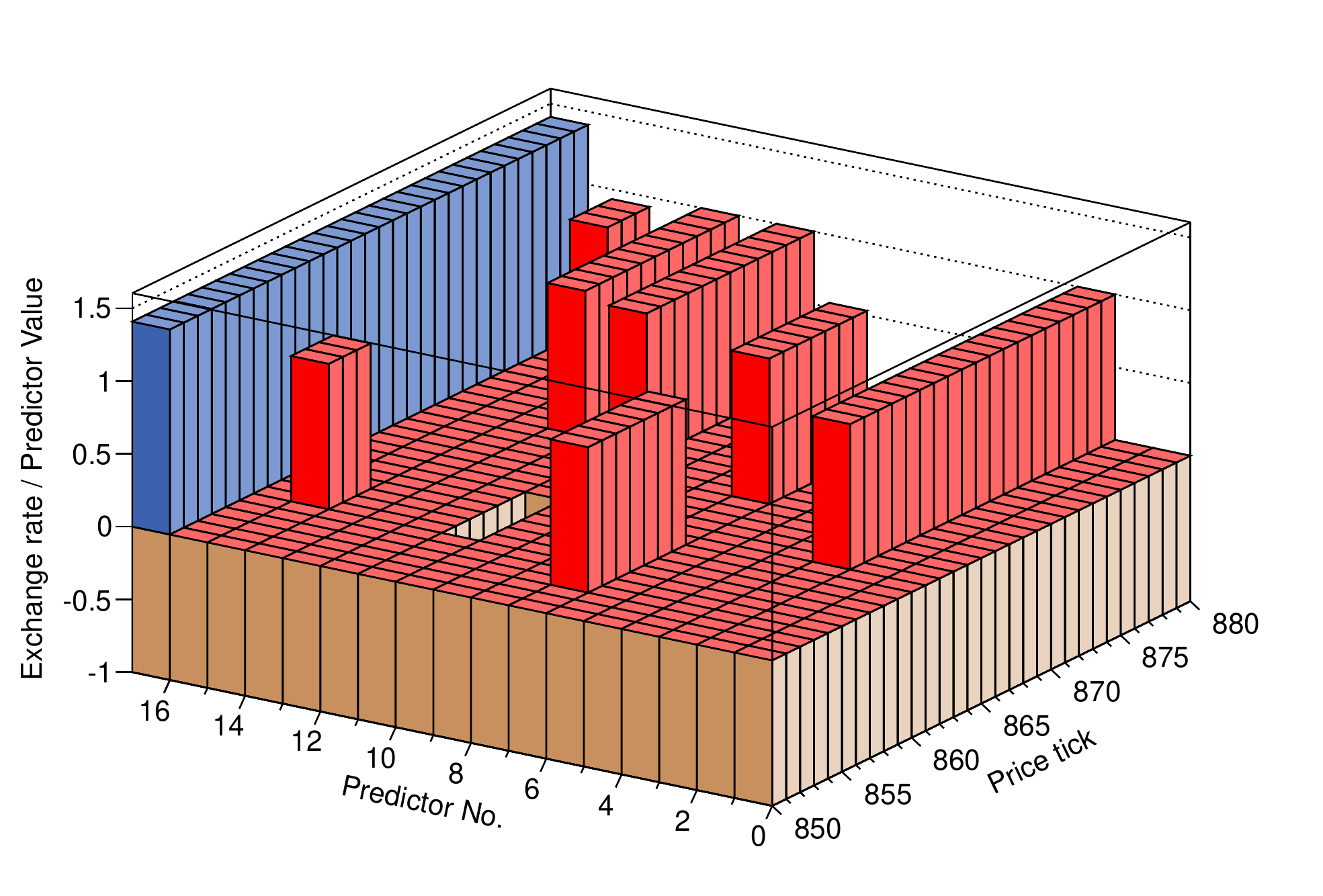}
	\label{fig:PredB}
}
\end{center}
\caption{The EUR/USD currency rate ticks from 2010/07/15 10:00:00 (blue dots) with the evaluated values of summarized prediction (red dots), normalize to $\pm1$ (a). The detail of subfigure (a) for a very short time period (b), the evaluated string momenta values (red) with the currency rate ticks on the background (blue).\label{fig:Pred}}
\end{figure}

The opening and closing of trade positions is determined by statistical evaluation of string momenta. In Fig.~\ref{fig:Pred}, such a statistical procedure is demonstrated graphically for $n_s=16$. Each new price tick leads to the evaluation of string momenta, in our case to sixteen values equal to $-1$, $0$, $1$. One can see the distribution of evaluated values in Fig.~\ref{fig:PredB} for a very short time period. A few positive values and one negative are clearly visible, the others are equal to zero. Then in Fig.~\ref{fig:PredB} the red dot represents the summarized value of the evaluated string momenta normalized to $\pm 1$. The blue dots are the EUR/USD currency rate ticks from 2010/07/15 10:00:00 up to the first $5\times 10^{5}$ ones. On the left of the subfigure one can see the instant of the ``learning'', when the statistics is gained and the momenta do not predict any values. Later the algorithm is fully operational.

\begin{figure}[!tb]
\begin{center}
\subfloat[]{
	\includegraphics[width=0.48\columnwidth]{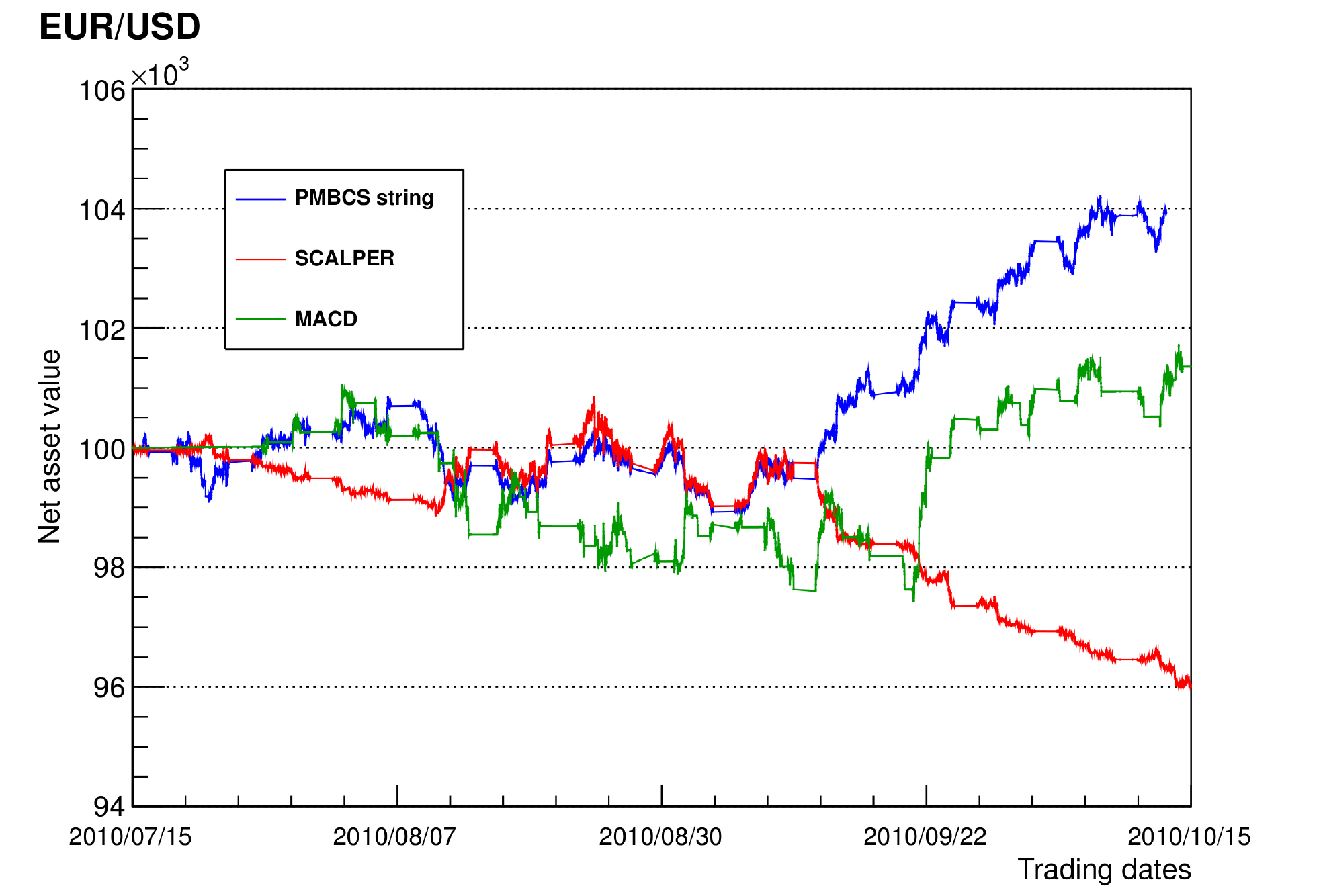}
	\label{fig:RegA}
}
\subfloat[]{
	\includegraphics[width=0.48\columnwidth]{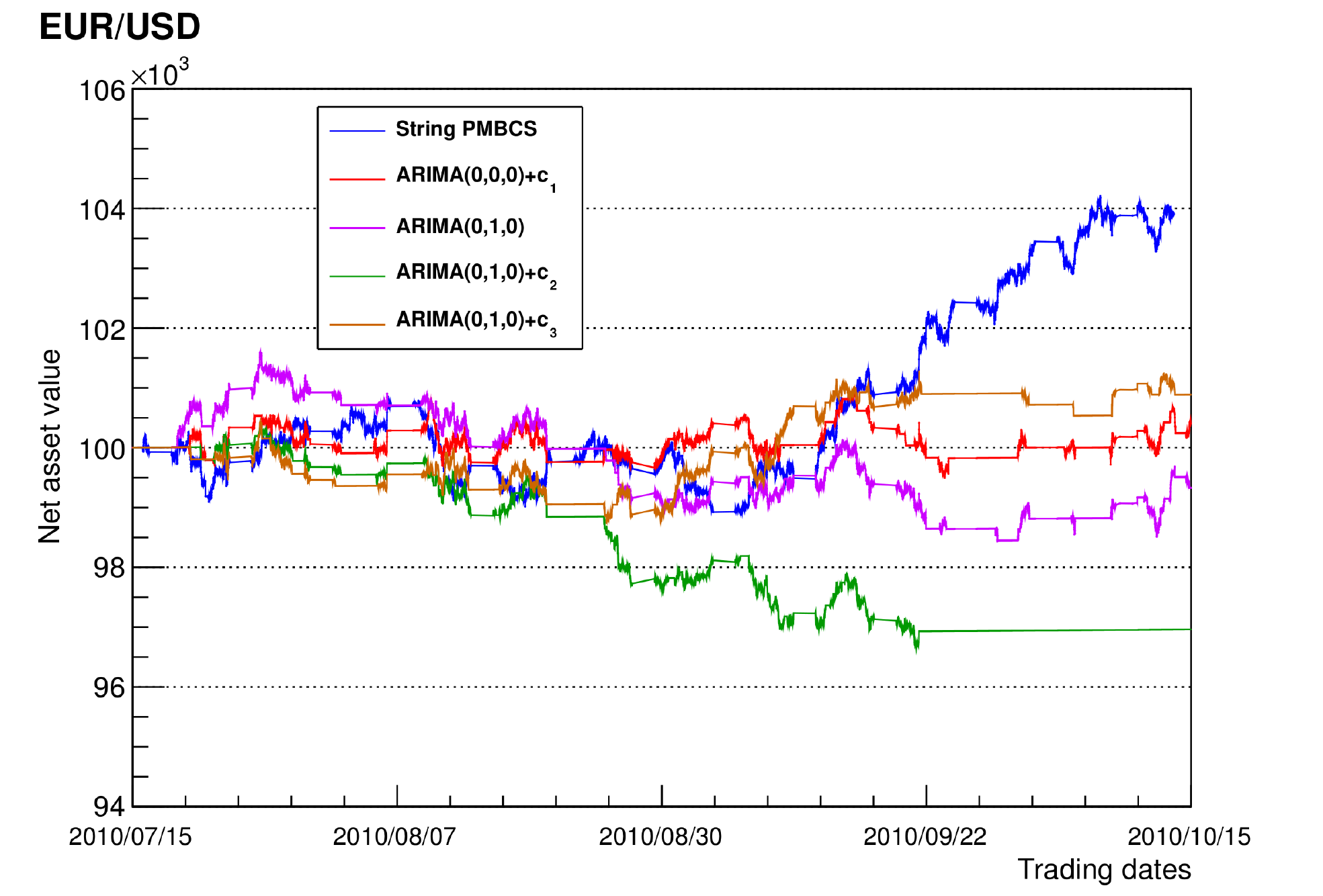}
	\label{fig:RegB}
}
\end{center}
\caption{The net asset value of the model on the EUR/USD currency rate for a selected time period compared with the values of basic time series forecasting models (ARIMA) and trading strategies (SCALPER, MACD) (see also Tab.~\ref{tab:Reg}). \label{fig:Reg}}
\end{figure}

\begin{figure}[!htb]
\begin{center}
\subfloat[]{
	\includegraphics[width=0.48\columnwidth]{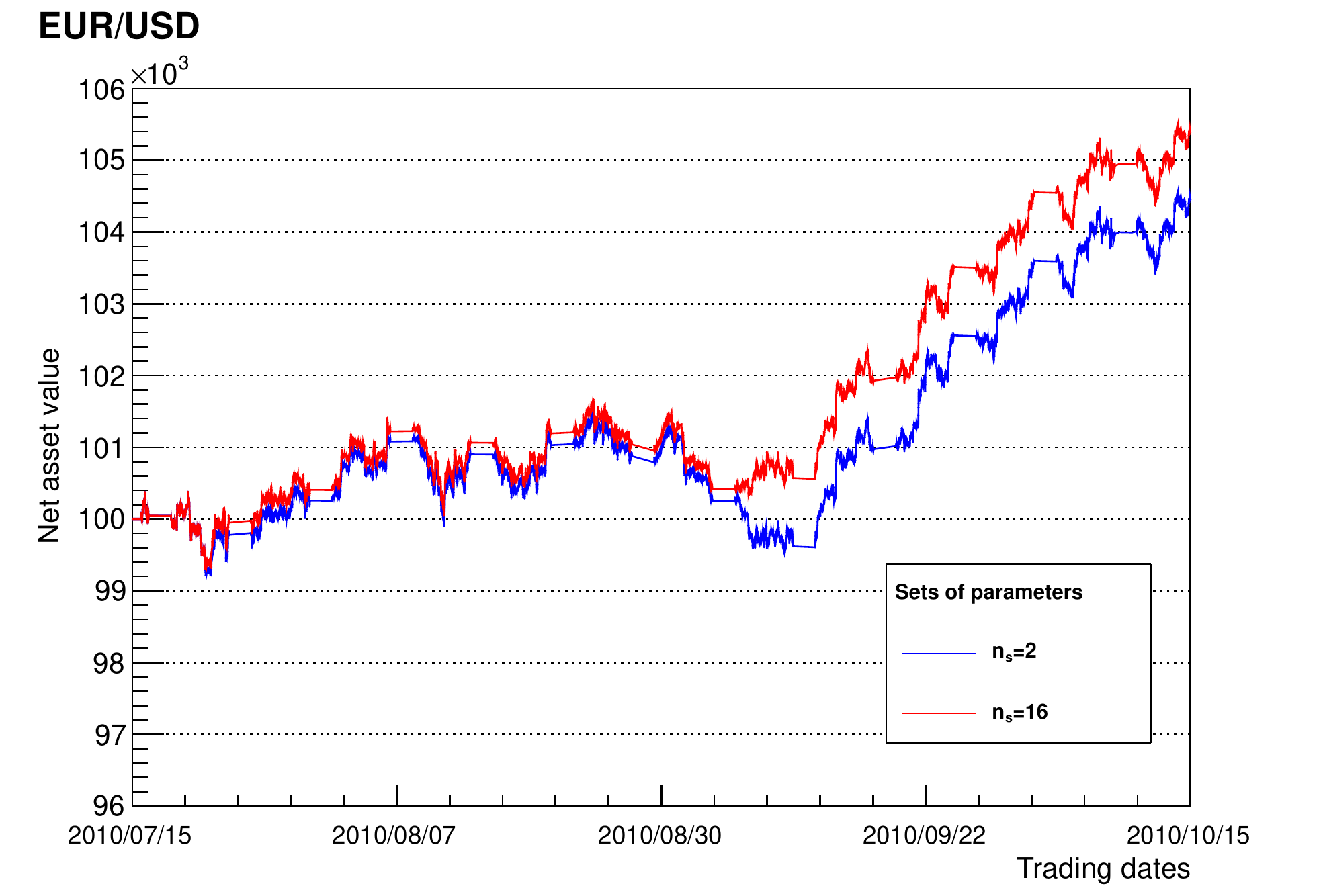}
	\label{fig:CompA}
}
\subfloat[]{
	\includegraphics[width=0.48\columnwidth]{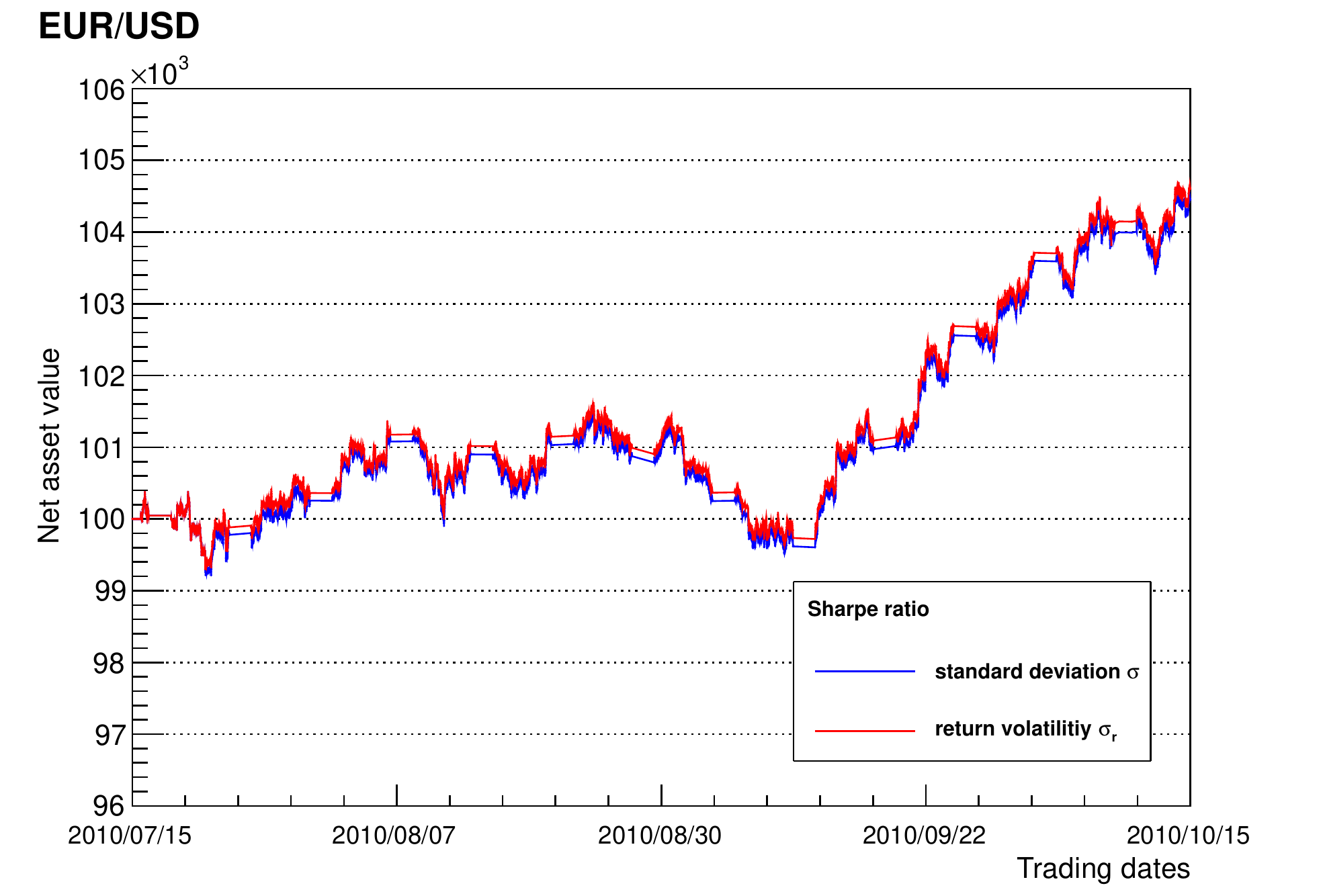}
	\label{fig:CompB}
}
\end{center}
\caption{The net asset value of the model on the EUR/USD currency rate for a selected time period as the dependence on the sets of trade parameters $n_s$ and as the dependence on the statistical methods (return volatility).\label{fig:funcComp}}
\end{figure}

The string PMBCS Self-learning model was benchmarked against the basic time series forecasting models and trading strategies. For this purpose we chosen the scalping strategy of taking profits on small price changes (SCALPER) \cite{ITS}, a trend-following momentum indicator Moving Average Convergence Divergence (MACD) based on the exponential moving averages (EMA) \cite{MACD} and finally the class of the autoregressive integrated moving average models (ARIMA) \cite{Arima, Arima2}, including ARIMA(0,0,0)$ +c$ – the mean constant model, ARIMA(0,1,0) – the random walk model and ARIMA(0,1,0)$ +c$ – the random walk with drift model (for different constants $c$). All mentioned models were implemented into the trade system as the corresponding algorithms. The results of the simulations are presented in Tab.~\ref{tab:Reg} and Fig.~\ref{fig:Reg}. As one can see the predicted NAV values after three months time period are close to zero profit for the most of the cases, especially for ARIMA models. Also the mean $\mu$ of NAV values oscillates around the starting point throughout the whole period. It is not surprising, one expects such behaviour for the random walk models where predicted values are equal to the last observed values. The reliability of the SCALPER method for examined period is also very small as it tends to the negative profit very rapidly. However, the scalping strategy is primary intended to take small profits for short time scales.

\begin{table*}[!ht]
\begin{center}
\begin{tabular}{>{\raggedright\hspace{2mm}}p{0.13\linewidth}>{\raggedleft}p{0.1\linewidth}>{\raggedleft}p{0.09\linewidth}>{\raggedleft\arraybackslash}p{0.1\linewidth}>{\raggedright\hspace{3mm}}p{0.18\linewidth}>{\raggedleft}p{0.1\linewidth}>{\raggedleft}p{0.1\linewidth}>{\raggedleft\arraybackslash}p{0.1\linewidth}}
\toprule
\textbf{Model} & \textbf{Mean \boldmath$\mu$} & \textbf{Sigma \boldmath$\sigma$} & \textbf{NAV [\%]} & \textbf{Model} & \textbf{Mean \boldmath$\mu$} & \textbf{Sigma \boldmath$\sigma$} & \textbf{NAV [\%]} \\
\midrule
String PMBCS  & $741$   & $1482$ & $4.33$  & ARIMA(0,0,0) + $c_1$      &    $95$ &  $249$ &  $0.40$  \\
SCALPER       & $-1396$ & $1340$ & $-3.99$ & ARIMA(0,1,0)              &  $-247$ &  $805$ & $-0.67$  \\
MACD          & $-383$  & $1020$ & $1.35$  & ARIMA(0,1,0) + $c_2$      & $-1699$ & $1208$ & $-3.04$  \\
              &         &        &         & ARIMA(0,1,0) + $c_3$      &    $44$ &  $679$ &  $0.89$  \\
\bottomrule
\end{tabular}
\caption{The comparison of the results for the net asset values (NAV) of our model and basic time series forecasting models (ARIMA) and trading strategies (SCALPER, MACD) on the EUR/USD currency rate for the time period 2010/07/15 – 2010/10/15. Mean $\mu$ is the average of the values (reference point $10^{5}$), $\sigma$ is the standard deviation and NAV is the percentage change of the start and end positions for the selected time period (see also Fig.~\ref{fig:Reg}).} \label{tab:Reg}
\end{center}
\end{table*}

\subsection{Volatility} \label{subsec:volat}

The risk of the algorithm is controlled by the trade strategy parameters, see Sec.~\ref{subsec:strategy}. Their precise determination is the matter of the empirical analysis and the setting of the degree of risk. However, we tried to find out how the Sharpe ratio (Eq.~\ref{eq:sr}) can influence the trading strategy. We focused on the volatility which refers to the standard deviation of currency returns of a financial instrument within a specific time horizon described by one half of the string length $l_s/2$ (see Sec.~6 in \cite{P:3}). The return volatility at the time scale $l_s/2$ is defined by
\begin{gather}
\sigma_r(l_s/2)= \sqrt{r_2(l_s/2)-r_1^2(l_s/2)},\qquad r_m(l_{\rm
s})=\sum_{h=1}^{l_s/2} \bigg(\frac{p(\tau+h)-p(\tau+h-1)}{p(\tau+h)}\bigg)^m,\qquad  m=1, 2.
\end{gather}

The brief result for the prediction dependence on the Sharpe ratio is presented in Fig.~\ref{fig:funcComp}. The set of string parameters was fixed at value $n_s=16$, the time period was from 2010/07/15 to 2010/10/15. The subfigure \ref{fig:CompA} describes the dependence on the standard sigma $\sigma$ as was always before, the subfigure \ref{fig:CompB} describes the influence of $\sigma_r$. The enhancement of the results is rather small, but it shows us that the investigation of Sharpe ratio dependencies must be taken into account seriously. The next improvement of the Self-learning model (by sophisticated evolutionary algorithms, statistical evaluation) could help to find optimal string parameters. On the other hand, the recent results, which did not choose the volatility of the market as a model indicator, e.~g., the methodology of scale of market quakes \cite{Bisig}, are quite promising and maybe the future version of the algorithm could show interesting progress in this field.

\section{Spin as a profit for long position} \label{sec:spin}

\begin{figure}[!tb]
\begin{center}
\subfloat[]{
	\includegraphics[width=0.48\columnwidth]{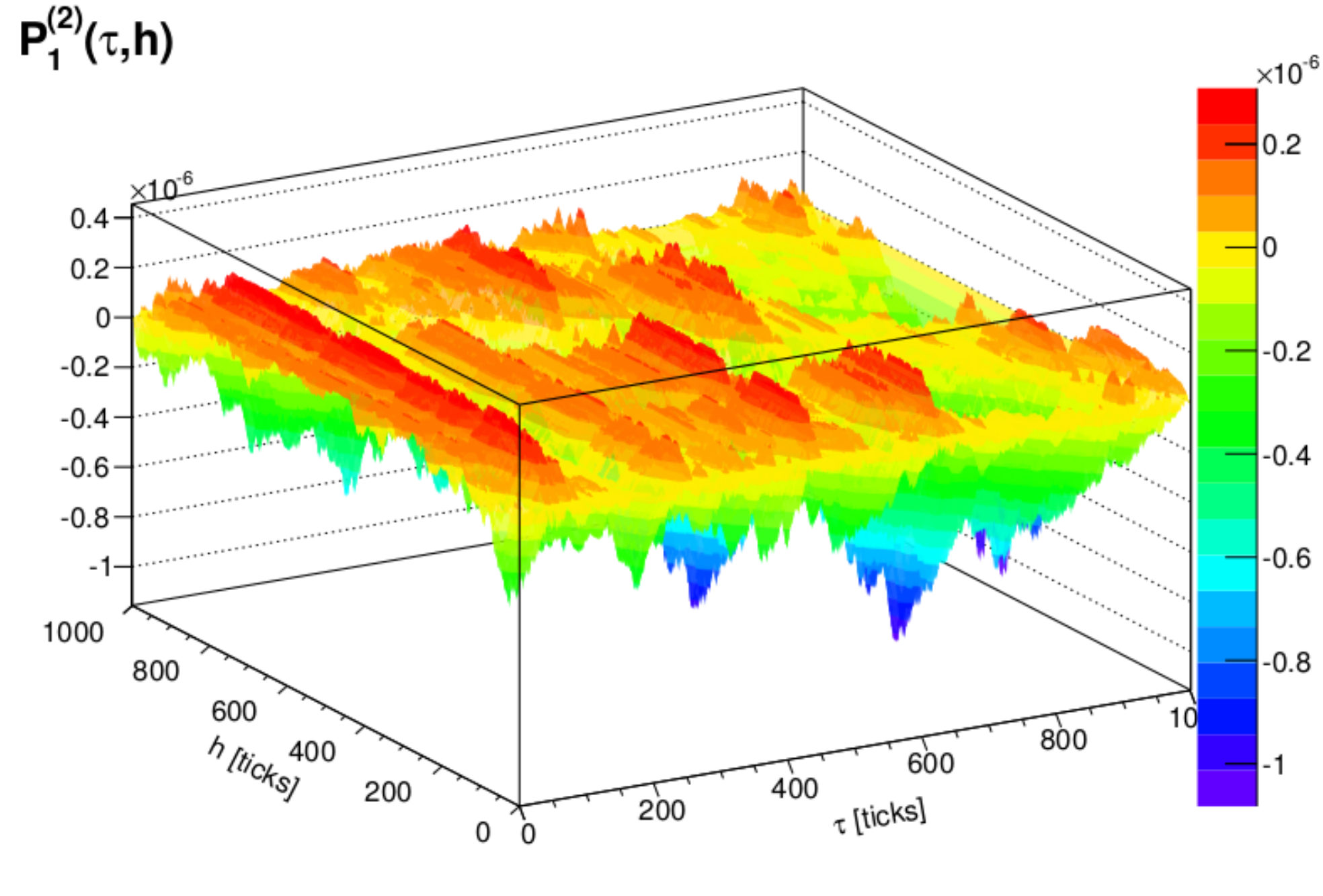}
	\label{fig:pxA}
}
\subfloat[]{
	\includegraphics[width=0.48\columnwidth]{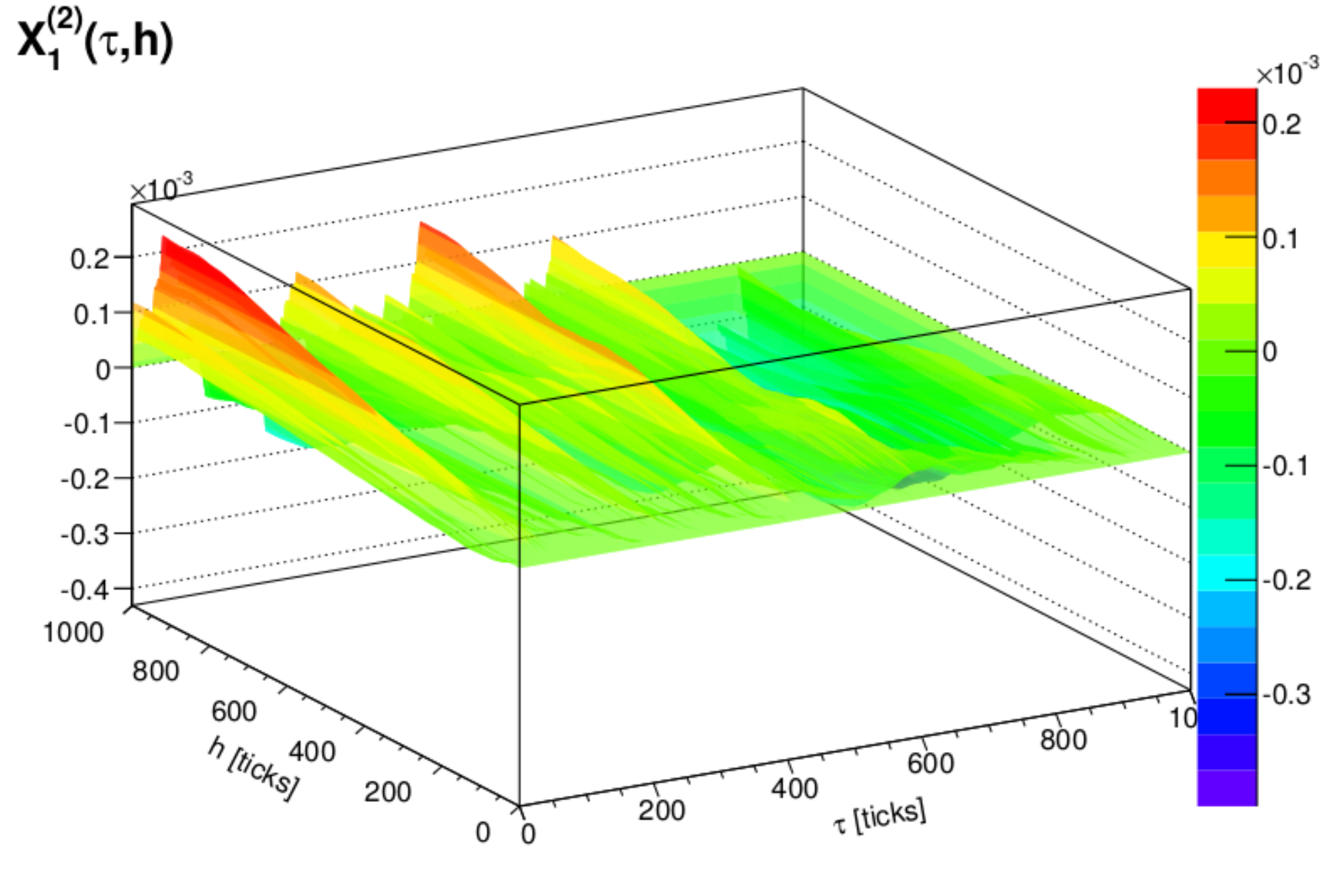}
	\label{fig:pxB}
}
\end{center}
\caption{The examples of the currency data map for EUR/USD OANDA data represented by the 2-end-point string map $P_{1}^{(2)}(\tau,h)$ (a) and the conjugate variable $X_{1}^{(2)}(\tau,h)$ (b). The calculation carried out for $l_s = 1000$, $q = 1$ at some time instant.\label{fig:px}}
\end{figure}

As another application of the string theory approach we would like to sketched the model based on the 2-end-point string maps $P_q^{(2)}(\tau,h)$ and their conjugate variables $X_q^{(2)}(\tau,h)$ (see Sec.~3 in \cite{P:3}). The choice for the variable $X$ is cleared from Fig.~\ref{fig:px}. In comparison to the 2-end-point string map $P$ its behaviour is smoother and hence it is less noisy. It highlights only the highest events in the market, so it seems to be the better object for the treatment of the statistics in our predictions.

Discrete dynamical rules are implemented where the string state is sequentially transferred to the past and stored by means of instant replicas. In this model, the $n$-th string of replica system is described by the tuple (for the simplification the indices ``$q$'' and ``(2)'' are omitted)
\begin{equation}
\left[\{X_{S}^{(n)}(\tau,h)\}; \, n=0,1,\ldots, N; \,h=0, \ldots,  h_{\rm op};\,
S^{(n)}\in  \{-1,1\}\, \right]
\end{equation}
including string coordinates and additional one spin supplementary variable $S^{(n)}$. The meaning of the spin is the same as in particle physics where there are two possibilities for the spin orientation of particle~\cite{spin}. Suppose the long position is opened at the quote $h_{\rm op}$ and closed at $h_{\rm cl}> h_{\rm op}$, then $S^{(n)}$ describes profit when $S^{(n)}=+1$ or loss when $S^{(n)}=-1$. In the case of buy order the sign of $S^{(n)}$ can be deduced from the price change according to $S^{(n)}=\mbox{sgn}(X_{\rm b} (h_{\rm cl}) - X_{\rm a}(h_{\rm op}))$.

The differences between string states can be measured by the string-string Hilbert $L_p$-distance as follows
\begin{equation}
D_p^{(n,N)} = \left[\frac{1}{d_{\rm X} h_{\rm op}} \sum_{h=0}^{h_{\rm op}}\sum_{j=0}^{d_{\rm X}} \Big| X_j^{(N)}(\tau,h) - X_j^{(n)}(\tau,h) \Big|^p\right]^{1/p}\,,
\end{equation}
where $n$, $N$ are the string-replica indices. The fuzzy character of the prediction of the spin variable of the $N$-th replica is described by means of
\begin{equation}
\overline{S}^{(N)} =
\sum_{n=0}^{N_{\rm red}} S^{(n)} w^{(n,N)}\,,
\quad
N_{\rm red} = N - ( h_{\rm cl} - h_{\rm op} ) ,
\end{equation}
which includes Boltzmann-like weights
\begin{equation}
w^{(n,N)} =\frac{\exp\left(
- c_{D} D_p^{(n,N)}/\overline{D}_p^{(N)}
\right)}{
\sum_{n'=0}^{N_{\rm red}}
\exp\left( - c_{D}\, D_p^{(n',N)}/\overline{D}_p^{(N)}\right)}\,,
\end{equation}
where the inter-replica distance is rescaled by the mean
\begin{equation}
\overline{D}_p^{(N)} = \frac{1}{N_{\rm red}+1}
\sum_{n=0}^{N_{\rm red}} D_p^{(n,N)}\,.
\end{equation}

The example of the distributions $S^{(n)}=+1$ and $S^{(n)}=-1$ as dependence on the interval $( h_{\rm cl} - h_{\rm op} )$ is represented by the histograms $h_{S+}$ and $h_{S-}$ in Fig.~\ref{fig:histmD} (both distributions are normalized to the interval $(0,1)$). In the case of $M$ distribution (Fig.~\ref{fig:histmC}) the relevant values of the mean lay in the close interval near $(0.3,0.4)$, i.~e., too low and high values of $M$ momenta were not important for the determination of Sharpe ratio in the Predictors Evaluator module. It seams that for the spin statistics is the situation opposite, the low values in the $S^{(n)}=+1$ distribution give the most predictive result. We hope that for the future analyses the spin statistics can bring more valuable results.

\subsection{Symbolic dynamics and inter-string information transfer}

We postulated dynamics as ordered moves of the data. The moves originate from the initial string $X_j^{(N)}(h)$ including transformation of data. The information then passes sequentially along copies $X_j^{(n)}$ in the sense of decremented replica index $n$ according to
\begin{eqnarray}
X_j^{(N)}(h)   & \leftarrow & X_j(h)\,, \quad S^{(n)} \leftarrow
\mbox{sgn}(X_{\rm b}( h_{\rm cl})-X_{\rm a}(h_{\rm op}))\nonumber,
\\
X_j^{(N-1)}(h) & \leftarrow & X_j^{(N)}(h)\,, \quad  S^{(N-1)}\leftarrow S^{(N)}\nonumber,
\\
&\ldots &\nonumber
\\
&\ldots &\nonumber
\\
X_j^{(1)}(h) &\leftarrow & X_j^{(2)}(h)\nonumber,
\qquad S^{(1)}\leftarrow S^{(2)}\nonumber,
\\
X_j^{(0)}(h)
&\leftarrow & X_j^{(1)}(h)\,, \qquad S^{(0)}\leftarrow
S^{(1)}\,.
\end{eqnarray}
We see that the information becomes lost at $X_j^{(n=0)}$. This method could be useful for trading algorithm especially for selection of final trades.

\section{Conclusion} \label{sec:con}

It is known by referring to the experimental data published in many journals that the random walk model is a good approximation of the market reality in static situations or in an equilibrium state of the financial market. Whenever such extra-ordinary events take place, especially leading to a financial or economical crisis, the random walk model fails. It means that in case of big instability with big volatility and fluctuation of prices, some other approaches need to be developed to describe and obtain real dynamics of such a market panic behaviour. This work is some attempt to explain and moreover predict financial data with a new string model approach.

Market equilibrium comes at the price of commodity for balancing the market forces like demand \& supply. In market equilibrium, the amount that the buyers want to buy is equal to the amount that the sellers want to sell. We call this equilibrium when the forces of demand \& supply are in balance, whereas there is no reason for a price to rise or fall as long as other factors remain unchanged. It is the situation in which the supply of an item is exactly equal to its demand. Since there is neither surplus nor shortage in the market, price tends to remain stable in this situation. Equilibrium price is also called market clearing price because at this price, the exact quantity that producers take to the market will be bought by consumers, and there will be nothing ‘left over’. For markets to work, an effective flow of information between a buyer and a seller is essential. The flow of information is described in our approach as an interaction part of the energy operator in kinetic and potential energy of the simulated strings.

This paper does not cover the full range of possible properties of the string theory approach. In addition to the theoretical modeling, the purpose of the paper was the study of the financial forecasting on the OANDA real data for the PMBCS model as well as the demonstration of some interesting properties of our model on real online trade system. As a result, the abilities of the self-learning model were shown to find the optimal string parameters for the final opening/closing of trade positions. The model was benchmarked against the most used class of the time series forecasting models and trading strategies, the simulations on OANDA data shows the higher values of the profit in comparison with the trend following strategies, e.~g. MACD, ARIMA. The stability of the algorithm on the transaction costs for long trade periods was demonstrated.
The next logical step is the improvement of the Self-learning algorithm for the data evaluation, as the received results are quite encouraging.

For another application of the string approach, we sketched some hierarchical model of algorithmic chemistry from string atoms to string molecules as a method of adaptive boosting. Discrete dynamical rules are implemented where the string state is sequentially transferred to the past and stored by means of instant replicas, as was developed in Sec.~\ref{sec:spin}. We defined a spin of strings which could detect a long-run profit where a fuzzy character of the prediction of the spin variable of the N-th replica can be investigated. Finally, inter-strings information transfer can be analyzed as an analogy with dynamic of prices or currency at specified exchange rate options.

And last but not least, the proper algebraic and geometric construction of the space of time series, possible under a Kolmogorov space concept with consistent separation axiom, could help with the analyses of nonstationary time series models, the volatility clustering phenomena in financial time series data or the separation of hidden Markov transition probability state in quantum entaglement state. There are some indications that such a concept could be realized as a topological space with a few hidden states in extradimensions of loop space of time series data. It will be interesting to join hidden states with the empiricaly observed characteristic correlation structures patterns (e.~g. \cite{MM}), but the other investigations must take place to the future.


\begin{acknowledgments}
The work was supported by the Science and Technology Assistance Agency under Contracts No. APVV-0171-10, No. APVV-0463-12, VEGA Grant No. 2/0037/13 and Ministry of Education Agency for Structural Funds of EU in the frame of the projects 26220120021, 26220120033 and 26110230061. R. Pincak would like to thank the TH division in CERN for hospitality.
\end{acknowledgments}

%
%

\end{document}